\documentclass[twocolumn,english,aps,pre]{revtex4-2}
\usepackage[T1]{fontenc}
\usepackage[utf8]{inputenc}
\usepackage{amsmath}
\usepackage{graphicx}
\usepackage{wasysym}
\usepackage{mathrsfs}
\usepackage{color}
\usepackage{latexsym}
\usepackage{hyperref}
\usepackage{subfigure}
\usepackage{wasysym}
\usepackage{bm}
\usepackage{float}
\usepackage{cancel}
\usepackage{comment}
\usepackage{soul}
\usepackage{subcaption} 
\usepackage{orcidlink}

\makeatletter

\usepackage{babel}

\makeatother

\usepackage{babel}
\begin{document}
  \title{Soft thermal diodes: grafted polymers provide a highly tunable thermal rectification }
  \author{Claudio Pastorino$^{\dagger,*}$ \orcidlink{0000-0002-4833-5999}}
  \email{claudiopastorino@cnea.gob.ar}
  \author{Alejandro Monastra$^{\natural}$}
  \author{M. Florencia Carusela$^{\natural}$\orcidlink{0000-0002-7200-8752}}
  \affiliation{$^{\dagger}$Departamento de Fí­sica de la Materia Condensada, Centro
  Atómico Constituyentes, CNEA, Av.Gral.~Paz 1499, 1650 Pcia.~de Buenos
  Aires, Argentina}
  \affiliation{$^{*}$Instituto de Nanociencia y Nanotecnologí­a, CONICET-CNEA, CAC.}
  \affiliation{$^{\natural}$Universidad Nacional de General Sarmiento; CONICET}
\begin{abstract}
We explore the heat-rectification properties of a two-phase fluid confined in a nano-chamber with one wall coated with end-grafted polymers using molecular-dynamics simulations. We find a significant thermal diode effect for a wide range of chamber fillings for both very stiff and fully flexible polymers. A stationary heat flux is imposed on the system by fixing the walls at two different temperatures, computing the heat flow as a function of the filling density. The fluid presents a liquid phase, located close to the cold wall, and, for many fillings, a vapor phase in contact with the hot wall. A vapor-liquid interface is also present and located at different distances from the walls, depending on the fluid filling. We study the system by comparing two operational modes: a direct mode, in which the polymers are grafted on the hot wall and exposed to the vapor phase. In the inverse mode, the wall temperatures are swapped and the fluid rearranges to adapt to the interchanged temperature gradient. We calculate  the mean heat flow, number density and temperature profiles for the stationary state in both modes. From them, we computed the density and temperature profiles, heat rectification coefficient, and resistivity profiles, for the two studied extreme cases of polymer bending rigidity. We found that,  the nano-chamber presents a significant degree of heat rectification with different characteristics and filling density ranges. We describe the conditions to be met across the chamber to obtain high thermal rectification as regards fluid filling, polymer properties and fluid-polymer affinity. These variables can be fine-tuned according to the application or material availability.  In the direction parallel to the walls the system is easily scalable towards macroscopic sizes, without affecting the thermal rectification. This  makes the soft-diode mechanism studied here very versatile in geometry, size and materials choice, which should lead to a wide field of applications.
\end{abstract}
\maketitle

\section{Introduction}
Over the last few decades, the increasingly intensive miniaturization and packaging of electronic devices have driven the need for innovations and improvements in small-scale thermal management solutions\cite{Hassan18}.
The substantial increase in power densities, as well as the presence of hot spots, can generate average heat fluxes exceeding 500-700 W/cm$^2$, which can result in excessive local temperatures, and seriously affect the performance of small devices in areas such as electronics, microfluidics, MEMS, wearables and biomedical devices\cite{Dhumal23,Smoyer19,Jung23,Jaya19,BarCohen21,Tachikawa22,Hassan18}.  In order to mitigate such issues, it is essential to implement effective thermal management strategies. These involve accurate control of temperature and the ability to drive heat flows in a preferential direction in order to efficiently dissipate its excess. The developed thermal architectures typically include thermal diffusers, active or passive thermal diodes, and thermal storage components\cite{Wehmeyer17, Cho19}.

Thermal diode are devices in which heat transport is rectified along a specific axis, that depends on the direction of the temperature gradient. They can be classified according to their structural and material features as either solid-state or soft (flexible) devices. The selection of a particular type depends on the specific application requirements. They can be composed of metals, ceramics, nano-structured materials, polymers, elastomers, surfactants, gels, and soft composite structures\cite{Pang24,Bianco22,JUN2025, Swoboda21,Wong21,Wehmeyer17, malik22,zurdo24}. 

The thermal rectification mechanism relies on structural asymmetries or on dynamical symmetry breaking, which can be achieved through different strategies:  Asymmetric Material Structures (mass-graded, geometrically asymmetric)\cite{wang2017, yousefi2020,carlomagno2020,Liu2019, zeng2008, zurdo2024}, interface-driven rectification\cite{lopez2018,xia2024, HAIYANGLI2024}, external-field modulation \cite{chen2021, hu2023}. In this scenario, (multi)Phase Change Materials (PCM) are excellent platforms for the design of thermal diodes. They consist of fluids or solids that can exhibit different phases simultaneously undergoing transitions in response to changes in temperature, exhibiting different thermal conductivities. In addition to the phase transition, it is necessary to introduce structural asymmetry in order to produce asymmetric thermal resistance for flux rectification.  Different strategies are found in the literature \cite{Wong21, Pang24,kommandur22}. In Refs. \cite{xia2024, JUN2025}, the authors proposed the addition of surfactants to the fluid. They found thermal rectification arising from an asymmetric surfactant adsorption on the walls when reversing the temperature gradient, producing an efficient thermal rectification ratio $R\sim 100$ for a considerable temperature difference.  Avanessian et al. \citep{Avanessian2016}  proposed a gas-filled chamber with heterogeneous surfaces to produce flux rectification. By non-equilibrium molecular-dynamics simulations the temperature differences and the solid-vapor interactions are tuned, finding a maximum rectification coefficient of $R\sim 7$. In another work \citep{AVANESSIAN18}, they added structural asymmetry in addition to the solid-vapor interactions, by introducing nano-pillars in one of the confining walls of the chamber. They obtained  a maximum rectification coefficient of $R\sim 100$. Other authors have proposed phase-change bridging-droplet thermal diodes with hydrophobic coatings, achieving a diodicity $\lesssim 100$ \citep{edalatpour20}.

This work presents an effective new strategy for thermal rectification that can be successfully applied to a wide range of polymer molecules, from soft to stiff, provided they are solvophilic. We study grafted polymers onto one of the diode walls, oriented perpendicularly to the direction of the temperature gradients. These coatings, combined with their interactions with the fluid (wettability), create an asymmetric thermal resistance. This facilitates the heat flow in a single direction, thereby promoting high heat-transfer efficiency and diodicity for a wide range of fluid fillings.

The structure of the text is as follows. In Section \ref{sec:Model}, we introduce the particle interaction model and the computational method. In Section \ref{sec:Results}, we present the thermal and rectification properties of a nano-chamber with a wall coated, either with very stiff, or fully flexible polymers. We also provide details of the fluid density and polymers' morphology for different chamber fillings. Finally, we present our conclusions and a final discussion in Section \ref{sec:Conclusions}.

\begin{figure*} 
\centering
\begin{tabular}{p{0.48\columnwidth}|p{0.48\columnwidth}}
\includegraphics[width=0.4378\columnwidth]{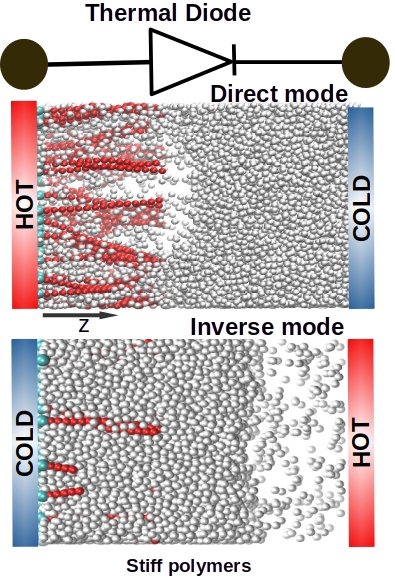} &
\includegraphics[width=0.4675\columnwidth]{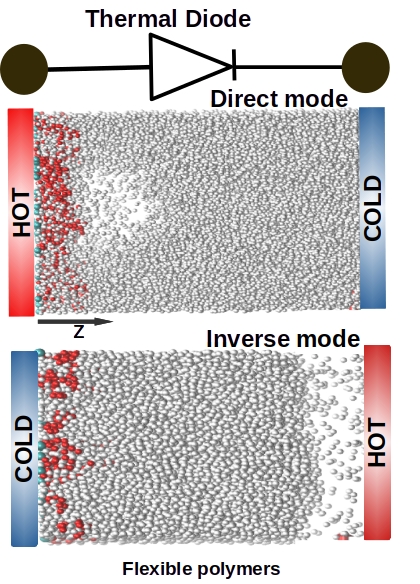} 
\end{tabular}
\caption{The top panel presents a sketch showing the thermal diode behavior relative to the nano-chamber. Sample configurations for the two states of the thermal diode with stiff (left panel) and fully flexible (right panel) polymers. In direct mode (DM) the polymers are grafted on the hot wall (upper row) and high thermal conductivity is found. Inverse mode (IM) corresponds to the case of  heat flowing from the  wall without polymers and usually presents lower thermal conductivity. 
In DM the polymers break up the vapor-liquid interface, producing liquid bridges (stiff polymers) or vapor bubbles (flexible polymers). The thermal conductivity is liquid-like, all across the channel. Inverse mode: The polymer-grafted wall is cold and the hot wall is exposed to a vapor layer. The thermal resistance is dominated by the vapor layer, close to the hot wall. \label{fig:system_sample} }
\end{figure*}

\section{Model and Computational Technique\protect\label{sec:Model}}
 We propose a soft thermal diode based on a nano-chamber made up of two walls confining a fluid,   which presents coexisting  liquid and  vapor phases. One of the walls is coated with  end-grafted polymers. The walls are kept at different temperatures, so that the system is subjected to a stationary temperature gradient (see sketch in Fig.\ref{fig:system_sample})  

We perform a comprehensive set of  molecular-dynamics simulations  to study the heat transfer, morphology and heat rectification properties as a function of fluid  filling in the chamber.   We use a coarse-grained model to account for all the molecular interactions, including polymers and fluid. We define a Lennard-Jones potential to account for the interaction among fluid particles, the non-bonded interactions of the polymers and the fluid-polymer interactions. The inter-particle truncated-and-shifted potential is given by
\begin{equation}
U(r)=U_{LJ}(r)-U_{LJ}(r_{c}),\,r<r_{c}\,,\label{eq:truncated_and_shifted_potential}
\end{equation}
and $U(r)$ vanishes for $r>r_{c}$. The cut-off distance $r_{c}$ depends
on particle types and $U_{LJ}(r)$ is given by:

\begin{equation}
U_{LJ}(r)=4\varepsilon_{\alpha\beta}\left[ \left( \frac{\sigma_{\alpha\beta}}{r} \right)^{12}-\left( \frac{\sigma_{\alpha\beta}}{r} \right)^{6}\right]\,,\label{eq:LJ_potential}
\end{equation}
where $\alpha=\mathrm{f,p}$ and $\beta=\mathrm{f,p}$ indicate the
particle types, fluid ($\mathrm{f}$) and polymer($\mathrm{p}$),
respectively. For the fluid particles, we have set $\varepsilon_{\mathrm{ff}} \equiv\varepsilon$,
$\sigma_{\mathrm{ff}}\equiv\sigma$ and $r_{c}^{\mathrm{(ff)}} = 2.5\sigma$. 
This Lennard-Jones cut-off has been studied thoroughly in the literature
and a wide knowledge of the phase diagram and interfacial properties
of the fluid has been reported \citep{Asano2012,Stephan_2018}. For
the polymer model, we use the Kremer-Grest coarse-grained
model\citep{Grest_86,Everaers2020}, with the addition of a bending
potential to account for the local stiffness of the chains\cite{Speyer_2019a,Speyer_17}. The Lennard-Jones
parameters have been set to $\varepsilon_{\mathrm{pp}}=\varepsilon$
and $\sigma_{\mathrm{pp}}=\sigma$ and $r_{c}^{\mathrm{(pp)}}=2.5\sigma$.
The fluid-polymer interaction has been set to produce a solvophylic behavior between polymer beads and fluid particles, with 
$\varepsilon_{\mathrm{fp}} = \varepsilon$  in Eq. (\ref{eq:LJ_potential}).
We also set $\sigma_{\mathrm{fp}}=\sigma$ and $r_{c}^{\mathrm{(fp)}} = 2.5\sigma$.
The mass of all particles and species is equal, and set to $m$.
In the coarse-grained Kremer-Grest
model \citep{Grest_86}, the polymer chain connectivity is described with a finitely extensible nonlinear elastic potential (FENE) of the form: 
\begin{equation}
U_{{\rm FENE}}=-\frac{1}{2}kR_{0}^{2}\ln\left[1-\left(\frac{r}{R_{0}}\right)^{2}\right]\,,\,r<R_{0}\,,
\end{equation}
(and $U_{{\rm FENE}}=\infty$ elsewhere), where the parameters were
set to the standard values $k = 30 \varepsilon/ \sigma^{2}$ and $R_{0} = 1.5 \sigma$.
The FENE potential is applied between consecutive monomers in a chain to account for its connectivity. We simulated 16-bead polymer chains.  Additionally, we define a bending potential, widely tested in previous works \citep{Speyer_2019a,Speyer_17,Speyer_15}, to model stiff polymers: 
\begin{equation}
U_{{\rm bend}}(\theta)=\frac{1}{2}k_{{\rm bend}}\theta^{2}\,,\label{eq:bend}
\end{equation}
where the bending constant $k_{{\rm bend}}$ controls the local rigidity
of the polymer, and $\theta$ is the angle defined by: 
\begin{equation}
\cos(\theta)=\frac{(\boldsymbol{r}_{i+1}-\boldsymbol{r}_{i})(\boldsymbol{r}_{i}-\boldsymbol{r}_{i-1})}{\left|\boldsymbol{r}_{i+1}-\boldsymbol{r}_{i}\right|\left|\boldsymbol{r}_{i}-\boldsymbol{r}_{i-1}\right|}\,,\label{eq:cos_theta}
\end{equation}
where $\boldsymbol{r}_{i-1}$, $\boldsymbol{r}_{i}$ and $\boldsymbol{r}_{i+1}$
are three consecutive connected polymer beads and $i=1,...,15$. This
potential form was also used to orient the polymers in the
direction perpendicular to the walls. An orientation force on bead
two is applied according to the equation (\ref{eq:bend}), and the angle
$\theta$ is calculated by setting $\boldsymbol{r}_{0}\equiv\boldsymbol{r}_{1}+\sigma\hat{z}$
in Eq. (\ref{eq:cos_theta}), where $\boldsymbol{r}_{1}$ is the grafted
head of the polymer (light blue particles in  Fig. \ref{fig:system_sample}). This choice means that very stiff polymers
have a strong tendency to orient perpendicularly to the grafting wall.

We consider two limiting cases of polymer stiffness: the case $k_{{\rm bend}} = 0$ (fully flexible
polymers), which reduces the polymer description to the original Kremer-Grest
model and $k_{{\rm bend}} = 256 \varepsilon$, which represents a very stiff polymer. The persistence length in the latter case is much longer than the contour length of the 16-bead polymers in the temperature range considered in this work.

The walls model of the nano-chamber serves two purposes. On one hand, we model the particle-wall interaction by a 9-3 Lennard-Jones potential, $U_{{\rm wall}}$, which only acts in the direction perpendicular to the walls: 
\begin{equation}
U_{{\rm wall}}(z)=A_{w}\left(\frac{\sigma_{w}}{z}\right)^{9}-\left|A_{w}\right|\left(\frac{\sigma_{w}}{z}\right)^{3}\,,\label{eq:wall_9_3_pot}
\end{equation}
with $A_{w}=3.2\varepsilon$ and $\sigma_{w}=\sigma$, where $z$
indicates the distance between each particle and the walls. These
parameters represent a moderately attractive, but impenetrable wall.
This potential mimics the van der Walls interaction between the particles
of the fluid or polymer and the wall particles, without including them
explicitly \citep{Mueller_00,Pastorino_06}. Additionally, we use the so-called
wall thermostat to set different temperatures on the top and bottom
walls in order to produce a stationary heat flux through the nano-chamber
and the liquid-vapor interface. We set a threshold of $z_{{\rm skin}}=1.2\sigma$.
When the particles have $z$ coordinate such that $\lvert z_{{\rm wall}}-z\rvert<z_{{\rm skin}}$
(wall zone), their velocities are overwritten by a random velocity
obtained from the distributions:

\begin{equation}
\rho_{x,y}=\sqrt{\frac{m}{2\pi k_{B}T_{w}}}\exp({-\frac{mu_{x,y}^{2}}{2k_{B}T_{w}}})\label{eq:dist_vels_gaus}
\end{equation}

\begin{equation}
\rho_{z}=\frac{m}{k_{B}T_{w}}\left|u_{z}\right|\exp({-\frac{mu_{z}^{2}}{2k_{B}T_{w}}})\,,\label{eq:dist_vels}
\end{equation}
where $u_{x}$, $u_{y}$ and $u_{z}$ are the Cartesian velocities
of the particles. $T_{w}$ indicates the wall temperature and $m$
the mass of the thermostated particles. Eq. \ref{eq:dist_vels_gaus}
is the Maxwell distribution, which is used for velocity components
tangential to the wall, and Eq. \ref{eq:dist_vels} is the probability
distribution for the velocity component normal to the wall. This last
equation samples the particles leaving the wall, i.e. the flux of particles
coming out from a wall, rather than the density of the particles in
the wall \citep{Tehver_98}. We used the ziggurat random number generator
for normal distributions in Eq. (\ref{eq:dist_vels}) \citep{Marsaglia2000}.
If the velocity of the particle points towards the interior of the
channel, the wall thermostat is not applied to the particle. We set
the hot and cold walls with the temperatures $T_{{\rm C}}=0.8\varepsilon/k_{B}$
and $T_{{\rm H }}=1.1\varepsilon/k_{B}$, respectively. This produces
a heat flux from the hot to the cold wall, through a vapor-liquid interface, as shown in Fig. (\ref{fig:system_sample}). This thermostat was widely used and tested in event-driven molecular dynamics simulations \citep{Tehver_98,Urrutia_14b,Paganini_15} and 
we have already applied it  with Lennard-Jones continuous potentials \citep{Pastorino2022}. 

The heat flux is measured directly in the simulations within the force routine in our molecular dynamics program. We implemented
the scheme described by Smith et al. \citep{Smith_2019} for a volume control given by a bin of width $1\sigma \times L_x\times L_y$ located close to the  center of the liquid phase of each simulation condition. We calculated the total heat flux as the average of kinetic (Eq. 24) and configurational (Eq. 25) contributions to the flux in Ref. \citep{Smith_2019}. The streaming velocity was set to zero because we do not have particle convection in the studied stationary states.

In this work, we use $\sigma$ as unit of length, $\varepsilon$ as unit of energy, $m$ as unit of mass, $\tau = \sigma \sqrt{m / \varepsilon}$ as unit of time, $\varepsilon / \tau $ as unit of heat flux, and $\varepsilon/k_{B}$ as unit of temperature. Considering argon as an example of a Lennard-Jones fluid \citep{Bugel2008}, these units would correspond to $\sigma$ = 3.408 \AA, $\varepsilon = 1.642 \cdot 10^{-21}$ J, $m = 6.634 \cdot 10^{-26}$ kg, $\tau = 2.166$ ps, $\varepsilon / \tau = 7.582 \cdot 10^{-10}$ W, and $\varepsilon/k_{B} = 118.9$ K.     

Firstly, we performed simulations in equilibrium conditions
by setting both walls to the same temperature  $T=0.8\varepsilon/k_{B}$,
with a filling $\rho_{f}$,  corresponding to the chamber approximately filled  with liquid and vapor and a  liquid-vapor interface close to the center of the chamber.

We performed extensive non-equilibrium molecular dynamics simulations for a nano-chamber of length $L = 40 \sigma$   in $\hat{z}$ direction with the confining walls kept at fixed and different temperatures, and varying the quantity of fluid inside the chamber. In all cases the simulation time step was $\delta t = 0.001 \tau $. The chamber walls have an area of  $900\sigma^2$ with lateral dimensions $L_x=L_y=30\sigma$. 
Furthermore, in one of the walls 16-bead polymer chains were  randomly end-grafted with a grafting density $\rho_g=0.071 \sigma^{-2}$ (number of grafted polymer end-beads per unit area). For a standard sample,  this is a total amount of  64 end-grafted polymers.
In the directions $\hat{x}$ and $\hat{y}$  we define periodic boundary conditions for the simulations. We
set the temperatures $T_{{\mathrm C}}=0.8\varepsilon/k_{B}$ and $T_{{\mathrm H}}=1.1\varepsilon/k_{B}$ on the cold and hot walls, respectively.  Observables are calculated after the system reaches the stationary state at time $t \approx 2000\tau$ ($2 \times 10^6$ MD steps), when a constant heat flux from the hot to the cold wall through the fluid is achieved. The physical quantities are measured in a standard trajectory of $180000 \tau$ ($18\times 10^6$ MD steps). 
The fluid forms a liquid-vapor interface, where the liquid phase is typically in contact with the cold
wall, while the vapor phase is exposed to the hot one. 
We quantify the fluid filling of the channel with the filling density
$\rho_{f}=N_{{\rm fluid}}/V$, where $N_{{\rm fluid}}$ is the total
number of fluid particles inside the chamber, independently if they
belong to the vapor or liquid phases.

\section{Results\protect\label{sec:Results}}

We studied the thermal properties
of the system as a function of filling density and upon swapping
of the hot and cold walls, to analyze the heat rectification capabilities
of the nano-chamber. We label as  {\sl direct mode} (DM) the condition in which  the grafted polymers are located  on the hot wall and {\sl inverse mode} (IM) the opposite heat flux direction. In the latter case, the polymers are grafted onto the cold wall. 
After reaching the stationary regime, the simulations were performed to compute all the properties in DM. Then, we swapped the wall temperatures and ran new simulations until the fluid and polymers rearranged themselves in the nano-chamber to accommodate the new thermal gradient. Once the system achieves the new stationary state (IM), the physical properties were calculated again.  
We present the results for  the thermal  and rectification properties of stiff polymers  ($k_{{\rm bend}}=256\varepsilon$ in Eq. \ref{eq:bend})
and flexible polymers ($k_{{\rm bend}}=0\varepsilon$ in Eq. \ref{eq:bend})  in subsections  \ref{subsec:stiff_polymers} and \ref{subsec:flex_polymers}, respectively. The nano-chamber behaves as a thermal diode when it presents heat flux asymmetry upon inversion of the heat flux direction.

\subsection{Heat rectification with stiff polymers\protect\label{subsec:stiff_polymers}}

\begin{figure}
\includegraphics[width=0.98\columnwidth]{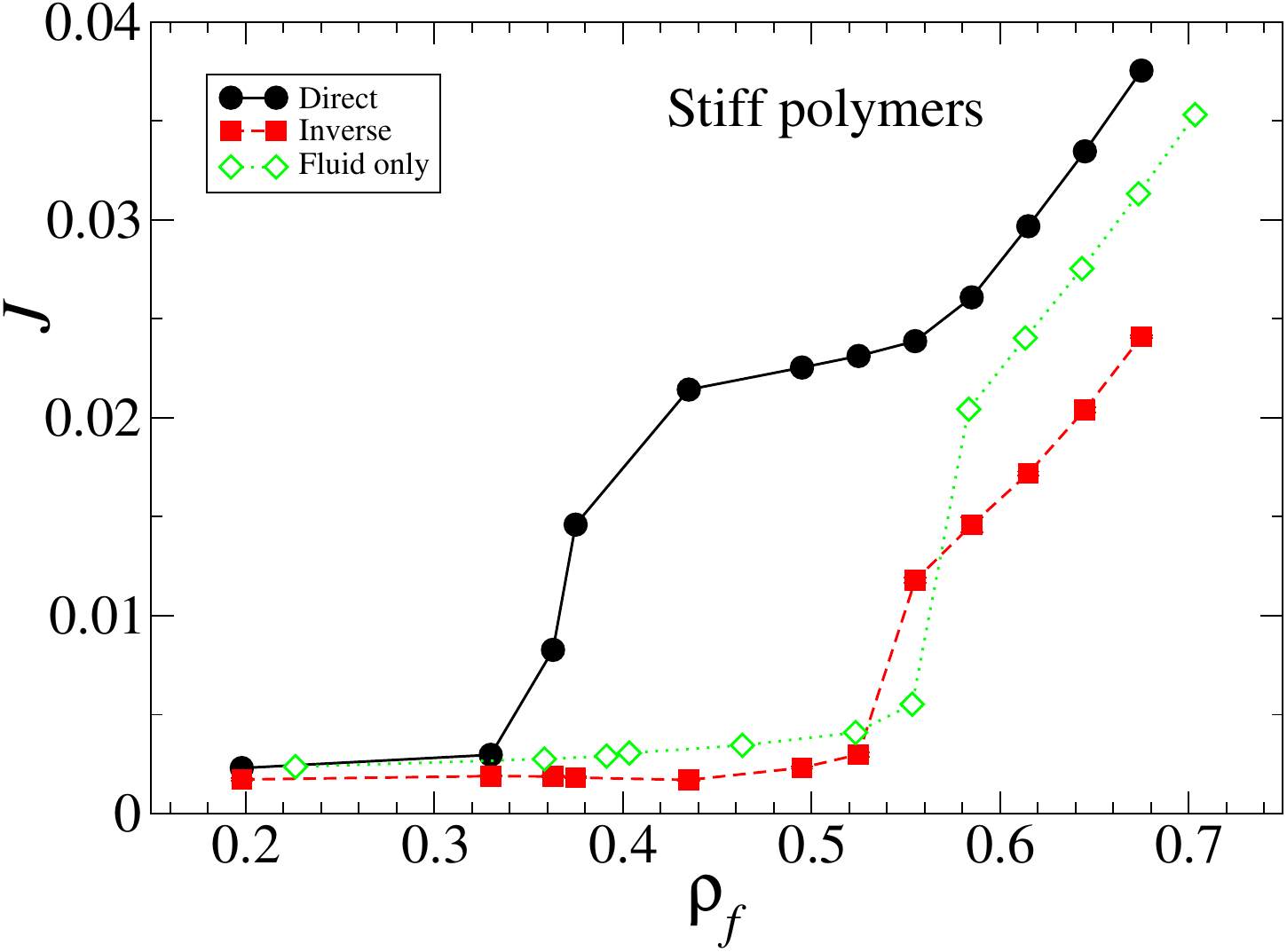}
\caption{
Heat flux (in units of $\varepsilon/\tau$) as a function of $\rho_f$ for chambers with stiff polymers coating on one wall. The black (circles) /red curve (squares) curves correspond to the chamber with the wall temperatures set in DM/IM respectively. The green curve (open romboids) corresponds to the heat flux in a chamber without grafted polymers. The lines are a guide for the eyes. The error bars are smaller than the symbol sizes.} \protect\label{fig:Heat-flux-stiff_polymers}
\end{figure}

In this subsection, we present the heat and structural properties of the  nano-chamber coated with very stiff polymers.   In this limiting case the polymers behave as stiff rods, whose Kuhn length is much larger than the contour length. 
The local structure of the chains on the coated wall is highly extended. This differs from the structure observed in flexible polymers exhibiting the mushroom-type regime of polymer brushes, as it will be shown in the next subsection.  The difference between the structures causes a significant change in the thermal properties, as well as a wider range of fillings showing thermal rectification, in comparison with flexible polymers.
In Figure \ref{fig:Heat-flux-stiff_polymers} we present the heat flux as a function of chamber filling $\rho_{f}$. In black curve (circles)
is shown the heat flux in DM and in red
(squares), the heat flux set in IM. 
To analyze the improvement of the thermal transfer due to the polymeric coating, we performed an independent set of simulations with bare walls in the chamber, shown in green.
DM, IM and bare walls cases present intervals of filling densities for which the heat flux is very low and grows with a small slope. But this interval is much narrower for DM, which starts to grow in a pronounced way at $\rho_f\gtrsim0.34\sigma^-3$.
Considering the region for which the vapor phase is
present, the interval $\rho_{f}\in[0.35\sigma^{-3},0.55\sigma^{-3}]$ shows the greater difference between the heat flux of DM and IM. Furthermore, the DM fluxes are always larger than the IM ones, even for small filling densities. This means that the presence of the stiff polymers have always an effect in the overall heat transfer through the nano-chamber, in spite of being much higher for intermediate fillings.

At low fillings ($\rho_{f}<0.35\sigma^{-3}$) the heat flux is very small, but in IM is slightly smaller. At higher fillings, when the vapor phase is almost absent and liquid-like densities of fluid wet both walls ($\rho_{f}>0.55\sigma^{-3}$), there is also a difference
in the heat fluxes, with the heat flux for DM significantly higher than for IM. 
On the other hand, for bare walls, the heat flux lies between the DM and IM values for most of the filling range.

In order to quantify the heat rectification,
we define the mean heat rectification coefficient as\citep{Avanessian2016}:
\begin{equation}
R_{\rm heat}=\frac{|J_{{\rm dir}}|-|J_{{\rm inv}}|}{|J_{{\rm inv}}|}\label{eq:rectification}
\end{equation}
where $J_{{\rm dir}}$ is the average heat flux in DM, which
is the direction of expected higher heat transfer and $J_{{\mathrm inv}}$ the mean heat flux in IM.
The factor $R_{\mathrm heat}$ quantifies the asymmetry of heat fluxes in both directions (DM and IM), normalized by the heat transfer in IM. 

\begin{figure}
\includegraphics[width=0.98\columnwidth]{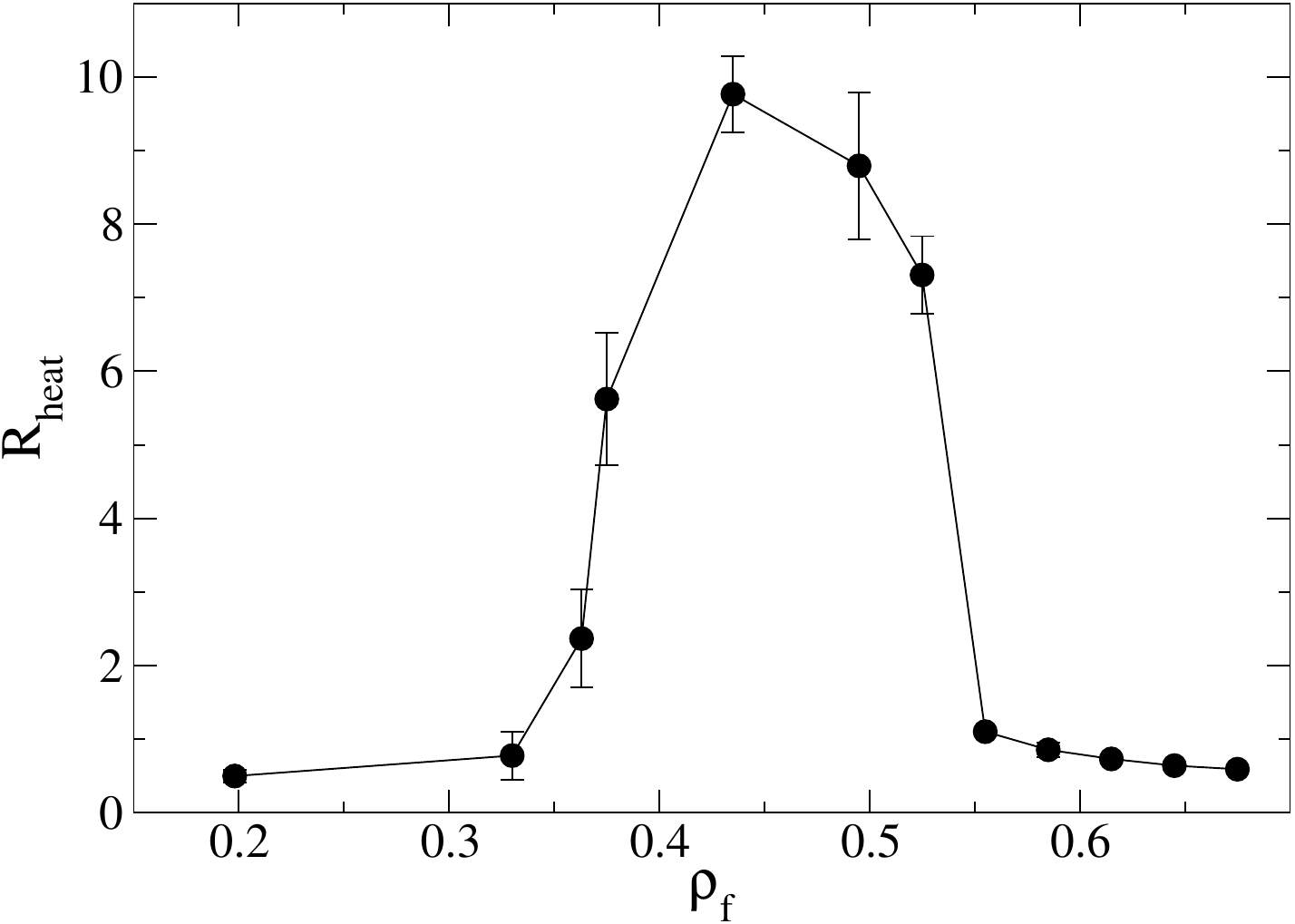}
\caption{Rectification factor  $R_{{\rm heat}}$ as a function of $\rho_{f}$ for nano-chambers coated with stiff polymers. The line is a guide for the eyes.}
\label{fig:Rectification-factor-stiff_polymers}
\end{figure}

In Fig. \ref{fig:Rectification-factor-stiff_polymers} we present $R_{{\rm heat}}$ as a function of 
$\rho_f$ for the nano-chamber coated with semiflexible (stiff) polymers. The grafting density was set $\rho_{g}=0.071\sigma^{-2}$ for all the cases and the polymers' end-beads were grafted randomly on one of the walls of the chamber. Non vanishing heat rectification is observed for the whole range of $\rho_f$. Moreover $R_{{\mathrm heat}}$ displays a non monotonic behavior with a maximum of $R_{\mathrm{heat}}=9.77(52)$ located at $\rho_{f}=0.436\sigma^{-3}$.
A wide interval of very high rectification is observed, however for $\rho_{f}>0.55\sigma^{-3}$, there is a steep reduction in $R_{{\rm heat}}$, which coincides with the disappearance of the vapor phase at such high chamber fillings. For practical applications, the interval 
$\rho_{f}\in [0.35\sigma^{-3},0.55\sigma^{-3}]$ is the most interesting, where high rectification is accompanied by a high heat flux in DM, as can be seen in Fig. \ref{fig:Heat-flux-stiff_polymers}.

In this regime, the substantial asymmetry in thermal flow is attributed to the following two phenomena. 
In the IM operational regime and for low filling, an homogeneous vapor phase is present between the hot wall and the liquid-vapor interface. However, in the DM operation, the stiff polymers arrange themselves in  tent-like structures with the neighboring chains. This brings them into contact with the liquid-vapor interface, favoring an increase in heat flux through the polymers. Additionally, this amplification is enhanced by the formation of ``liquid bridges'' in the surroundings of the polymeric-tent structures (see Fig. \ref{fig:system_sample}, left upper panel). 

As these tents are pretty rigid and stable, they provide a favorable physical-chemical environment for the fluid particles, which are effectively absorbed by the polymers towards the hot wall. This combination of liquid bridges plus polymer structures (tents) provides a very good route for heat transfer, as compared to the vapor phase. 

We argue that the system is scalable in the lateral directions ($\hat{x}$ and $\hat{y}$), without changing the thermal behavior in the heat flow direction ($\hat{z}$) i.e.\, keeping the heat rectification, that we find for a small sample. The reason for this is that the partial wetting produced by the polymers and the tent-like structures (morphology) of the chains  are independent of 
lateral area of the sample ($L_x \times L_y$), providing that the grafting density $\rho_g$ is kept constant. We tested this rationale by simulating a 
sample of lateral area four times larger ($L_x=L_y=60\sigma$) at the filling density of  $\rho_f=0.436\sigma^{-3}$, corresponding to the peak of rectification factor for stiff polymers. We obtained a value of $R_\mathrm{heat}=8.92(42)$, which presents no significative differences with the value obtained from the smaller original sample.

\begin{figure}
\includegraphics[width=0.80\columnwidth]{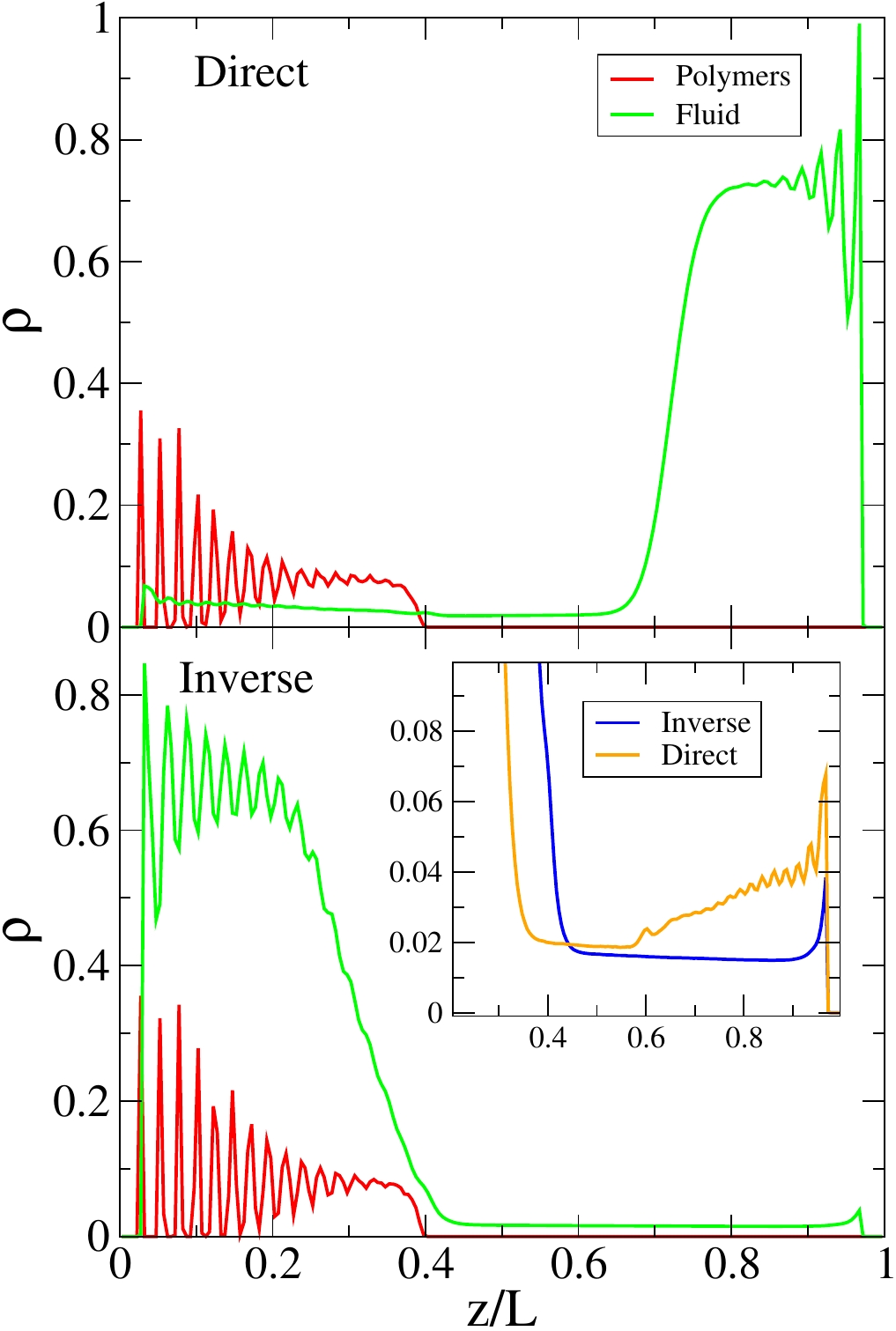}

\caption{Density profile of the channel coated with stiff polymers in direct
(upper panel) and inverse modes (lower panel). The fluid with a filling
density $\rho_{f}=0.198\sigma^{-3}$ is shown with green lines and
the density of stiff polymers is shown in red. The inset shows a detail for the fluid density in both direct (orange) and inverse (blue) modes locating, for comparison, the cold wall at $z/L=0$ and the hot wall at $z/L=1$\protect\label{fig:Density-profile-stiff_pol_low_filling}}
\end{figure}

\begin{figure}
\includegraphics[width=0.80\columnwidth]{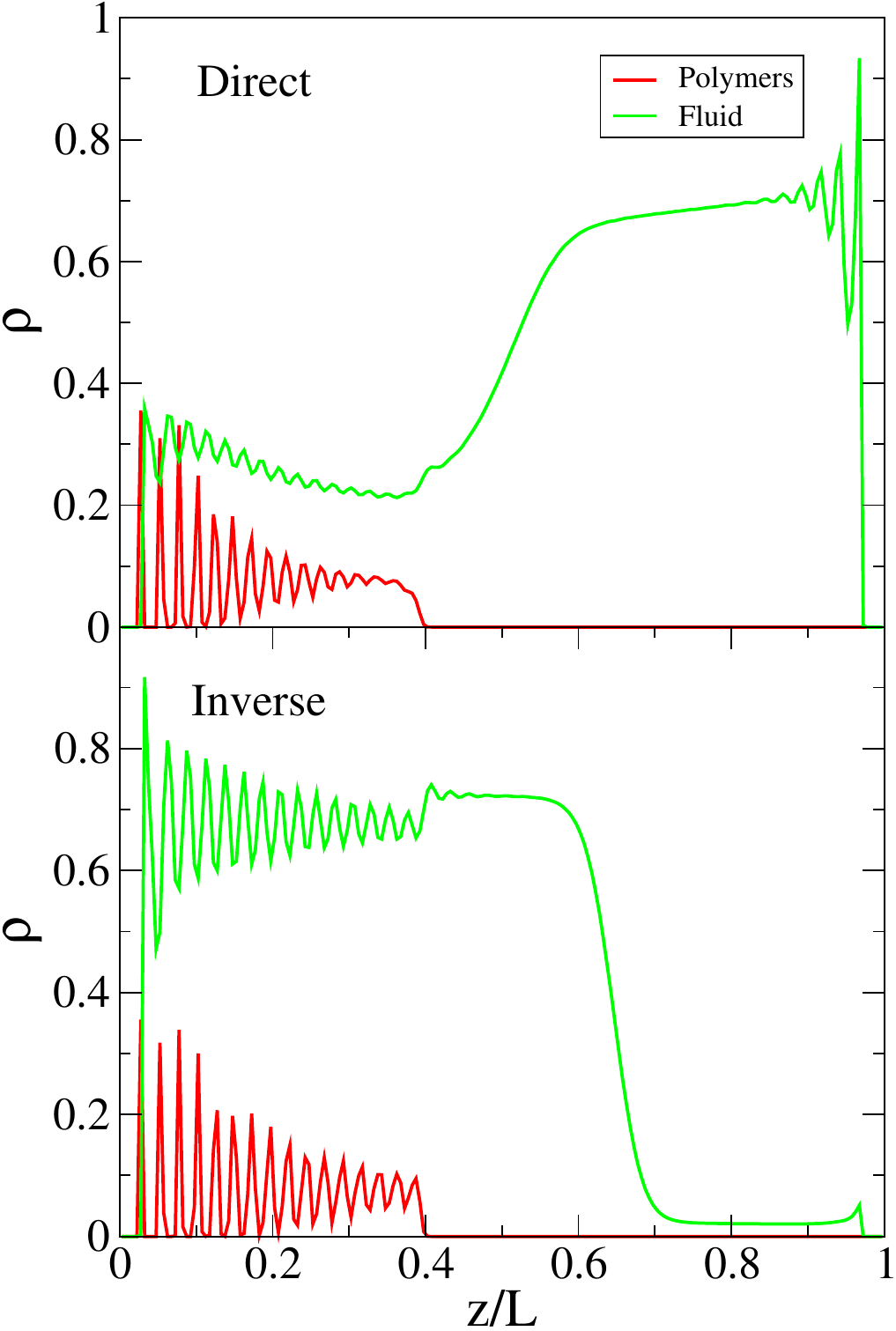} 

\caption{Density profile of the channel in direct (upper panel) and inverse
modes (lower panel) for a fluid filling of $\rho_{f}=0.437\sigma^{-3}$. Fluid
is indicated with green and the density of grafted polymers is shown
in red lines. The high fluid density is observed throughout
the whole channel, giving rise to a large heat flux, which is not
modified significantly by the presence of the grafted polymers on
the hot or cold walls.\protect\label{fig:Density-profile_stiff_polymers_middle_filling}}
\end{figure}

\begin{figure}
\includegraphics[width=0.80\columnwidth]{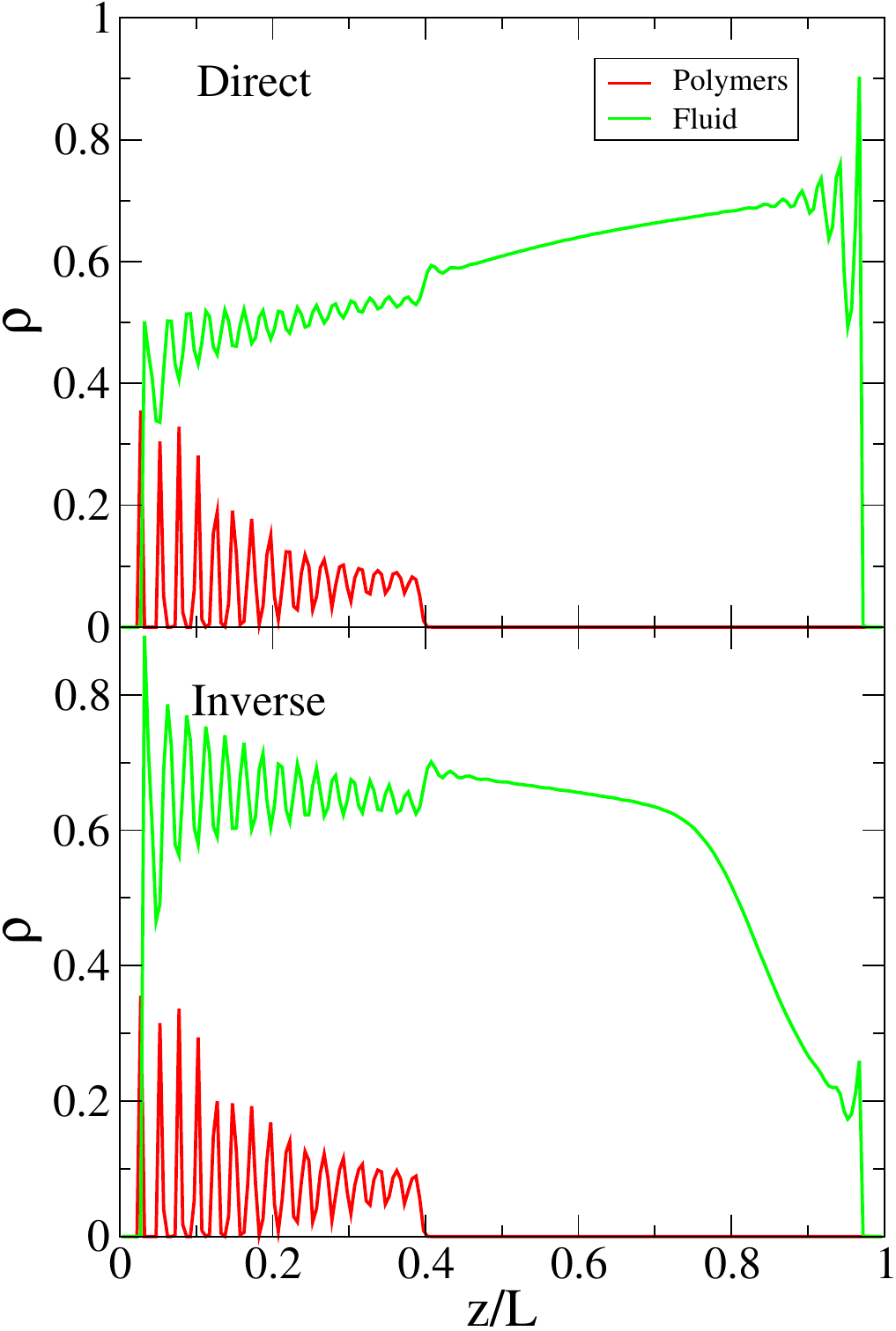}

\caption{Density profile of the channel in direct (upper panel) and inverse
modes (lower panel) for a fluid filling of $\rho_{f}=0.585\sigma^{-3}$.
Fluid is indicated with red and the grafted polymers' density is shown
in red curves. The high fluid density is observed throughout
the whole channel, giving rise to a large heat flux, which is not
modified significantly by the presence of the grafted polymers in
the hot or cold walls.\protect\label{fig:Density-profile_stiff_polymers_higg_filling}}
\end{figure}

To provide insight into the fluid distribution inside the nano-chamber, we show the density profiles of fluid (green curves) and polymers
(red curves) for three representative fillings which display distinctive features
of thermal behavior (Figures \ref{fig:Density-profile-stiff_pol_low_filling}
to \ref{fig:Density-profile_stiff_polymers_higg_filling}). In Figure
\ref{fig:Density-profile-stiff_pol_low_filling} we show the density
profile for a low fluid density ($\rho_{f}=0.198\sigma^{-3}$) in DM (upper panel) and IM (lower panel).
The green curve show the typical density of liquid phase ($z\lesssim14\sigma$),
with a region with layering caused by the local order in $z$ induced
by the hard wall (upper panel, Fig. \ref{fig:Density-profile-stiff_pol_low_filling}.
Afterwards, a vapor-liquid interface, followed by a vapor phase ($z\apprge15\sigma)$
which extends towards the hot wall. The polymer layer grafted on the
left wall, which is the hotter in the upper panel and the colder in
the lower panel shows a very ordered structure in $z$ which is very
similar either if the liquid wets the bare wall (upper panel) or the
polymer-coated wall (lower panel). This strong order is revealed by
the strong acute peaks in the density, which signals the monomers
positions almost fixed in $z$ coordinate. 
This is a marked contrast with the case of flexible polymers (see, for example, Fig. S1 in Supplementary Material). In that case, the polymers  show a much blurred density, which is also much closer to the grafting wall.
The very high stiffness
of the polymers, make them extend towards the center of the channel
to a distance practically equivalent to that of the contour length
of the chains. At this low fluid density the vapor phase is wide, dominant and present in both modes. However,
as shown in the inset of Fig. \ref{fig:Density-profile-stiff_pol_low_filling},
the vapor density is not the same. The inset shows inverse
(blue lines) and direct (orange line) modes in a reference frame with
cold wall located in $z=0\sigma$, for both cases. In a significant
part of the vapor phase, the density of the direct mode is higher
than that of the inverse mode and also it presents a positive slope,
with density enhancement towards the hot wall. This is counter intuitive
because the density grows towards higher temperatures, instead of
decreasing which is the usual case, as the particles reduce kinetic
energy, minimizing energy at lower temperatures. This can be attributed
to the presence of the extended polymers grafted on the hot wall for
the case of direct mode. They provide a favorable physico-chemical
environment as compared to the empty space ($z/L\in[0.375\sigma,0.55\sigma]$),
due to the attraction between monomers and fluid particles. The polymers
provide an extended attractive zone for the fluid particles. This
effect is not present in the inverse mode, in which the liquid wets
the polymer at the cold wall, and the vapor extends in a long region,
with negative slope towards the hot wall (see blue curve in the inset
of Fig. \ref{fig:Density-profile-stiff_pol_low_filling}) . 

In Figure \ref{fig:Density-profile_stiff_polymers_middle_filling}
we provide the densities of polymers and fluid for an intermediate
channel filling of $\rho_{f}=0.437\sigma^{-3}$, for which the rectification
is maximal, corresponding to the peak
of Fig. \ref{fig:Rectification-factor-stiff_polymers}. The main mechanism
of heat rectification is revealed in the density profiles. In inverse
mode (lower panel) the liquid phase and the polymers are both placed
at the cold wall, and an extended vapor phase is present towards the
hot wall (region of $z/L\gtrsim 0.75$). This mode presents, therefore, 
a zone of high resistivity due to the existence of a homogeneous vapor region. In contrast,
in direct mode (upper panel, Fig. \ref{fig:Density-profile_stiff_polymers_middle_filling} ) the interaction of the stiff polymers
with the fluid induce  liquid bridges\citep{Pastorino2022}, which
have liquid-like thermal behavior (high thermal conductivity). The vapor region
which would exist without grafted polymers, is now transformed by a coexistence
of liquid threads with fluid wetting the polymers and vapor bubbles (see left upper panel of Fig. \ref{fig:system_sample}.
The heat is transferred  mainly through the liquid bridges, giving rise to an overall resistance  much lower than in the inverse
mode. For this density, the rectification is at its peak with the
highest difference between the direct and inverse modes. 

In Figure \ref{fig:Density-profile_stiff_polymers_higg_filling}
the density profiles of a nano-chamber at high filling ($\rho_{f}=0.585\sigma^{{{}^-3}}$)  is shown.
The fluid density is high enough such that there is no vapor phase in the chamber,
but the profiles are anyway asymmetric. In direct mode (upper panel)
the fluid density slowly decays towards the hot left wall, in which the fluid wets
the polymer. The layering peaks of the polymer density, due to its high stiffness
(red curve) has  concomitant  deeps in the liquid density. The overall thermal behavior is of liquid-like density (and
heat resistance). In inverse mode, high-density liquid and polymers
are located at the the cold wall, giving rise to a total high density
there, followed by a significant reduction of the liquid density
in the neighborhood of the hot wall. This arrangement of fluid gives
rise to a slightly higher resistance in that zone, as we will describe later. 

We calculated the heat resistivity profile across the nano-chamber, by
the independent computation of the temperature profile and the heat
flux\citep{Pastorino2022}. A heat resistivity profile, $R(z)$ can
be derived from a local application of the Fourier equation to obtain:
\begin{equation}
R(z)=-\frac{1}{J_{z}}\frac{\partial T(z)}{\partial z},\label{eq:fourier-resistance}
\end{equation}

where $J_{z}$ is the mean heat flux and the gradient of the temperature
profile is obtained by deriving numerically the temperature profiles
$T(z)$ obtained in the simulations. The  temperature profiles from which the temperature gradient is calculated are 
presented in Fig. S2 of Supplementary Material.
In Figure \ref{fig:Resistance-profiles_stiff_polymers} we present
the resistivity profiles corresponding to the density profiles shown
in Figs. \ref{fig:Density-profile-stiff_pol_low_filling}-\ref{fig:Density-profile_stiff_polymers_higg_filling}
to relate the fluid and polymer density across  the nano-chamber with
the thermal properties. The resistivity profiles are shown with the
cold wall located at $z/L=0$ and the hot wall at $z/L=1$
for DM (black lines) and IM (red lines). Close to both walls, we observe peaks in the
resistivity.
These are present for all the studied  cases and they  are signaling the Kapitza resistance of the interfaces between the 
polymer and/or fluid with the hard walls. 

In the upper panel we observe the low-filling density case and the
DM (black curve) and IM (red curve) are quite different in
the vapor-phase region and they are  very similar in the liquid region ($z/L<0.25$).
As a general feature, the resistivity is higher in the inverse case
through most of the vapor phase region of the chamber. This gives rise to a higher
thermal resistance of the nano-chamber in IM, as compared
to DM. The total resistance of the chamb can be thought
as the integral over the resistivity profile, and therefore it is
clear that the resistance in inverse mode is higher than that of direct
mode, due mainly to  differences in the resistivity of the vapor-phase zone.
We note that even when the difference in DM and IM  behaviors is clearly  resolved,  there are considerable fluctuations of the resistivity profile for IM (red curve) in the vapor zone ($z/L>0.5)$. This is due to a low density of particles in the vapor phase which gives rise to a much larger statistical inaccuracy, as compared to the liquid phase. These fluctuations are  present even when we extended the simulation time to $400\times 10^6$ time steps, for the cases with a significant vapor phase in IM (Top and Middle panels of Fig. \ref{fig:Resistance-profiles_stiff_polymers} ).   
For both DM and IM, the Kapitza resistance peak, corresponding
to the liquid-vapor interface, is observed in the interval $z/L\in[0.25,0.5]$.
In DM there is an additional peak at $z\sim0.425\sigma$,
which we attribute to another interfacial zone between the vapor and the
polymeric phase (see upper panel, black curve). After that, the resistance
has a negative slope, consistent with the increase in fluid density
shown in the inset of Figure \ref{fig:Density-profile-stiff_pol_low_filling}.
In IM (red curve) the opposite behavior is observed: an increase
in the resistance toward the hot wall. This is also consistent with
the inset (Fig. \ref{fig:Density-profile-stiff_pol_low_filling}),
which shows a decreasing vapor density  in IM, i.e when there are not polymers grafted into the hot wall.
The shift in the Kapitza peak of IM, as compared to DM
is also expected, due to the shift of the interface position, observed
also in the inset of fluid densities. The polymers, wetting the liquid
in the cold wall, contribute to an overall increase of particles in
the liquid-like region of the nano-chamber, pushing forward the liquid-vapor
interface. In the middle panel of Figure. \ref{fig:Resistance-profiles_stiff_polymers}
the heat resistivity profile is shown for the intermediate filling density
of $\rho_{f}=0.437\sigma^{-3}$ , for which the rectification factor is at the maximum. The huge asymmetry in resistivity for both modes
is very clear: In direct mode (black line), the resistivity is very
low and practically liquid-like all along the nano-chamber. In IM
(red curve), we see the behavior of a liquid-vapor interface.
There is a zone of very low liquid-like resistivity up to $z/L\sim0.625$,
followed by a much higher resistivity ($z/L\gtrsim0.625$), responsible
for the much higher total thermal resistance in IM.
In this zone, the Kapitza peak of the liquid-vapor interfacial
resistance is observed at $z\sim0.71\sigma$, followed by a zone of
very high resistivity, corresponding to the vapor phase. Then, towards
$z/L\sim 1$, the peak corresponding the vapor-solid interface
is observed. This is  very similar to the case of lowest filling (upper panel,
Fig. \ref{fig:Resistance-profiles_stiff_polymers}). Finally, we show the resistivity profile for the regime of high filling, with $\rho_{f}=0.555\sigma^{-3}$  in the lower panel of Fig. \ref{fig:Resistance-profiles_stiff_polymers}.
For this case, there is no vapor interface and the fluid fills the
chamber with a relatively high, liquid-like density. This filling density is very similar to that shown in Fig. \ref{fig:Density-profile_stiff_polymers_higg_filling}, to show fluid and polymers densitiy profiles.
The Kapitza resistance of the liquid-vapor interface is therefore
absent and only the interfacial resistivity peaks with the solid walls are present, in the vicinity of the confining wall, i.e. at $z/L\sim0$ and $z/L\sim 1$. However, the resistivity profiles in DM and IM,  still show an asymmetry, depending on to which walls the polymers are grafted. In DM (black line), close to the hot
wall, the thermal resistivity is slightly lower as compared with IM (red line). This difference is coherent with the dip in the
fluid density observed for IM (see lower panel of Fig.
\ref{fig:Density-profile_stiff_polymers_higg_filling}), which is
not present in DM. In the latter the polymers grafted on
the hot wall help to form a much higher fluid density in that zone of the nano-chamber.
We also think that the high stiffness of the polymers provide a channel of  
high thermal conductivity through the polymer backbones, which is larger than the conductivity through the fluid. 
It has been reported in the literature that the heat conductivity of a single  polymer chain 
can be very high and much higher than that of a polymer melt\citep{Henry_08, Henry2014,Liu2012}. This remains to be studied thoroughly for the nano-chamber with end-grafted polymers, by computing separately
the heat flux contributions of the polymers and  the fluid. 

The resistivity profiles studied here complete the characterization
of the thermal properties and deepen the understanding of the asymmetry
of heat transfer for the different filling regimes of the nano-chamber
coated with stiff polymers. These are are fully consistent with the
density profiles, the heat flux and the rectification factors presented
before. 

\begin{figure}
\includegraphics[width=0.75\columnwidth]{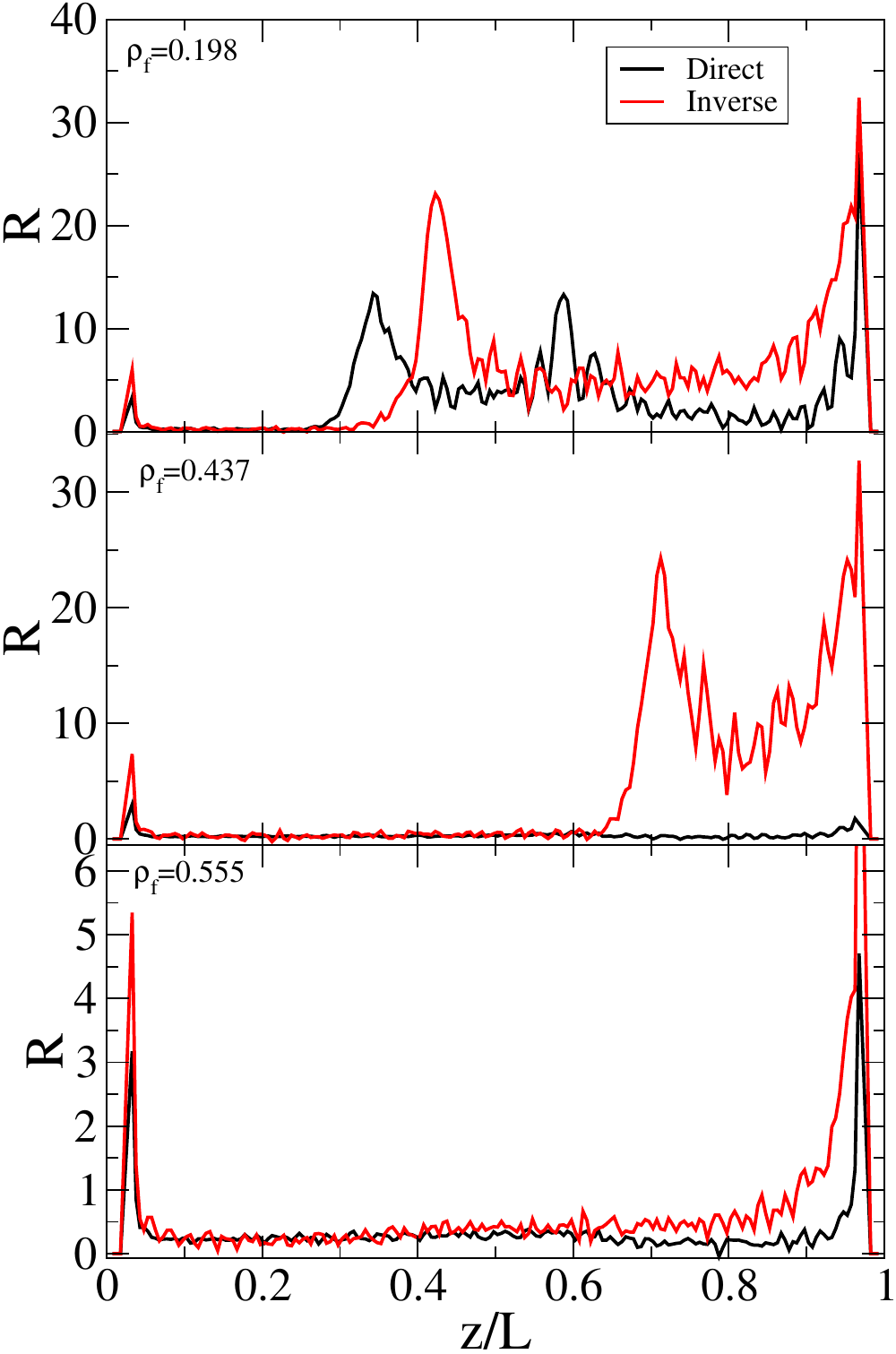}

\caption{Resistivity profiles for three important cases cases of stiff polymers:
The lowest studied fluid density $\rho_{f}=0.198\sigma^{-3}$ (upper panel), the density that gives
rise to the higher rectification factor (center panel, $\rho_{f}=0.437\sigma^{-3}$)
and a high fluid density in which the vapor phase vanishes (lower
panel, $\rho_{f}=0.555\sigma^{-3})$. The profiles were adjusted to
have the cold wall at $z/L=0$ and the hot wall at $z/L=1$,
in both, DM (black lines) and IM (red lines). \protect\label{fig:Resistance-profiles_stiff_polymers} }
\end{figure}

\subsection{Heat rectification with flexible polymers\protect\label{subsec:flex_polymers}}

\begin{figure}
\includegraphics[width=0.98\columnwidth]{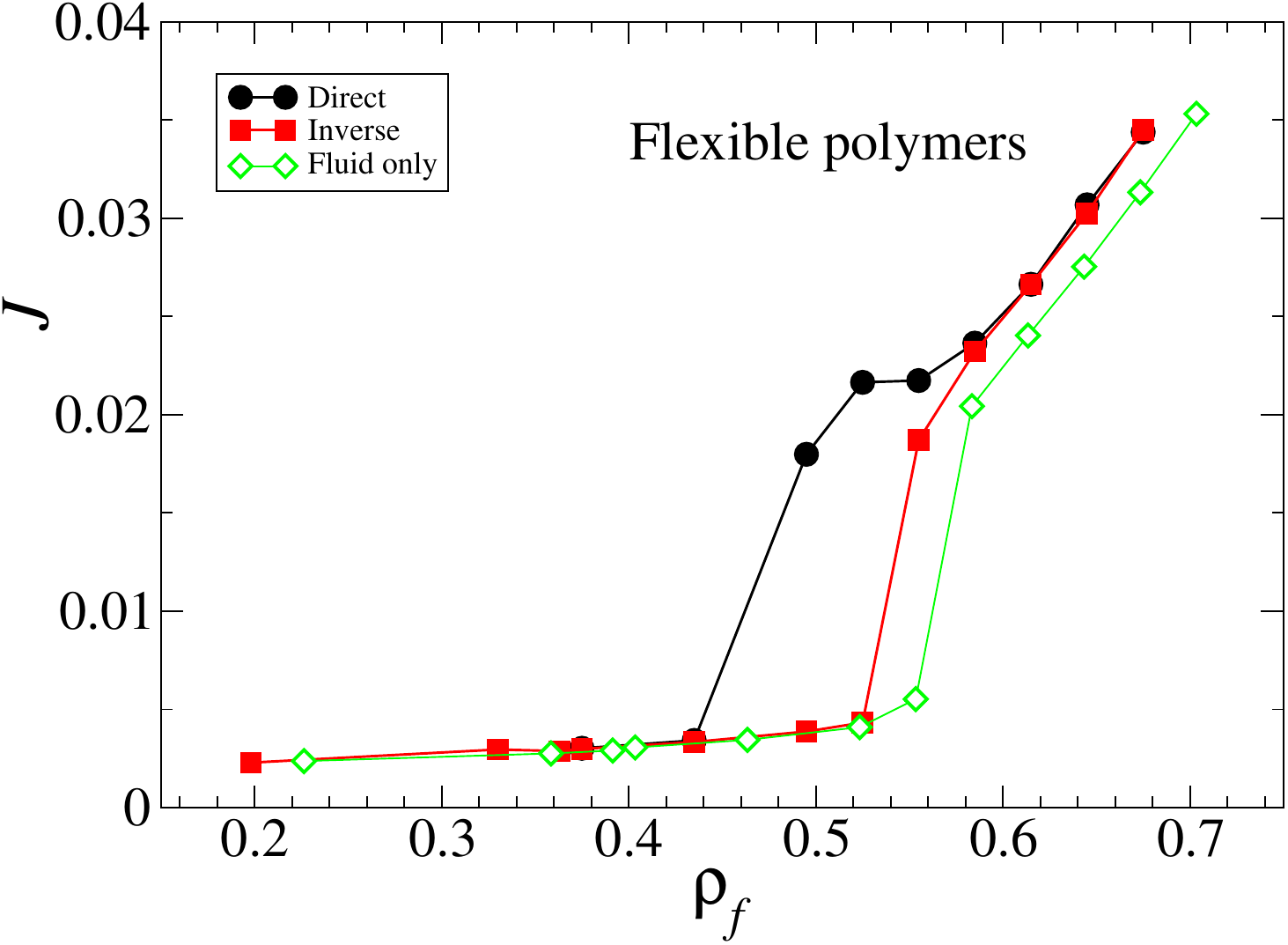}\caption{Heat flux as a function of channel filling for flexible polymers.
We compare the behavior for the channel with the polymers grafted
on the hot wall (DM, black curve, circles) and with polymers
on the cold wall (IM, red curve, squares). In green channel
(open diamonds) with exactly the same fluid filling but without polymers
on the walls is shown for comparison. The error bars are smaller than the size of the symbols. For those fillings in which the heat flux in direct and inverse modes differ, the setup presents rectification. \protect\label{fig:Heat-flux-flex_pols}}
\end{figure}

We performed another set of molecular-dynamics simulations for  a model chamber with one of the walls coated by fully flexible polymers end-grafted randomly, with the same grafting density than that used for stiff polymers ($\rho_{g}=0.071\sigma^{-2}$). We calculated the mean heat flux \textit{J} in the stationary regime\cite{Smith_2019}, which is shown in 
Figure \ref{fig:Heat-flux-flex_pols},  as a function of chamber filling. 
 The DM, in  black line and circles, is obtained for the polymers
grafted on the hot wall. The red line and square symbols shows the heat
flux for the same cases of $\rho_f$  with the wall temperatures
swapped (IM).  The green line and diamonds shows the heat flux ($\textit{J}$) versus
$\rho_{f}$ for a chamber without polymers, presenting only fluid
particles confined between bare walls. The three curves display a common general behavior: for
low  $\rho_{f}$ there is a small heat flux which increases slowly upon
filling. In this case, heat transfer is dominated by the presence of the vapor phase, as it was also observed for stiff polymers. Then a jump in the heat flux is observed for the three curves, but it happens at different values of $\rho_{f}$.
This difference in the heat fluxes for DM or IM in the range $\rho_{f}\in[0.4\sigma^{-3},\,0.58\sigma^{-3}]$ is  responsible of the  heat rectification, which is also observed for flexible polymers.
For $\rho_{f}\geq0.58\sigma^{-3}$ (high $\rho_{f}$), $\textit{J}$ is high and increases linearly. For these cases the liquid wets
both the hot and cold walls and the vapor phase vanishes. In this
regime there is no rectification, and the heat flux is approximately
the same for DM and IM. The heat flux is also slightly
higher as compared with a chamber filled only with fluid (green line).
We attribute this difference between the channel with and without
polymers, to an enhancement of heat transfer through the polymer beads.
The chain connectivity produces an effectively smaller mean free
path among particles exchanging kinetic energy, as compared with unconnected
liquid particles. 

\begin{figure}
\includegraphics[width=0.9\columnwidth]{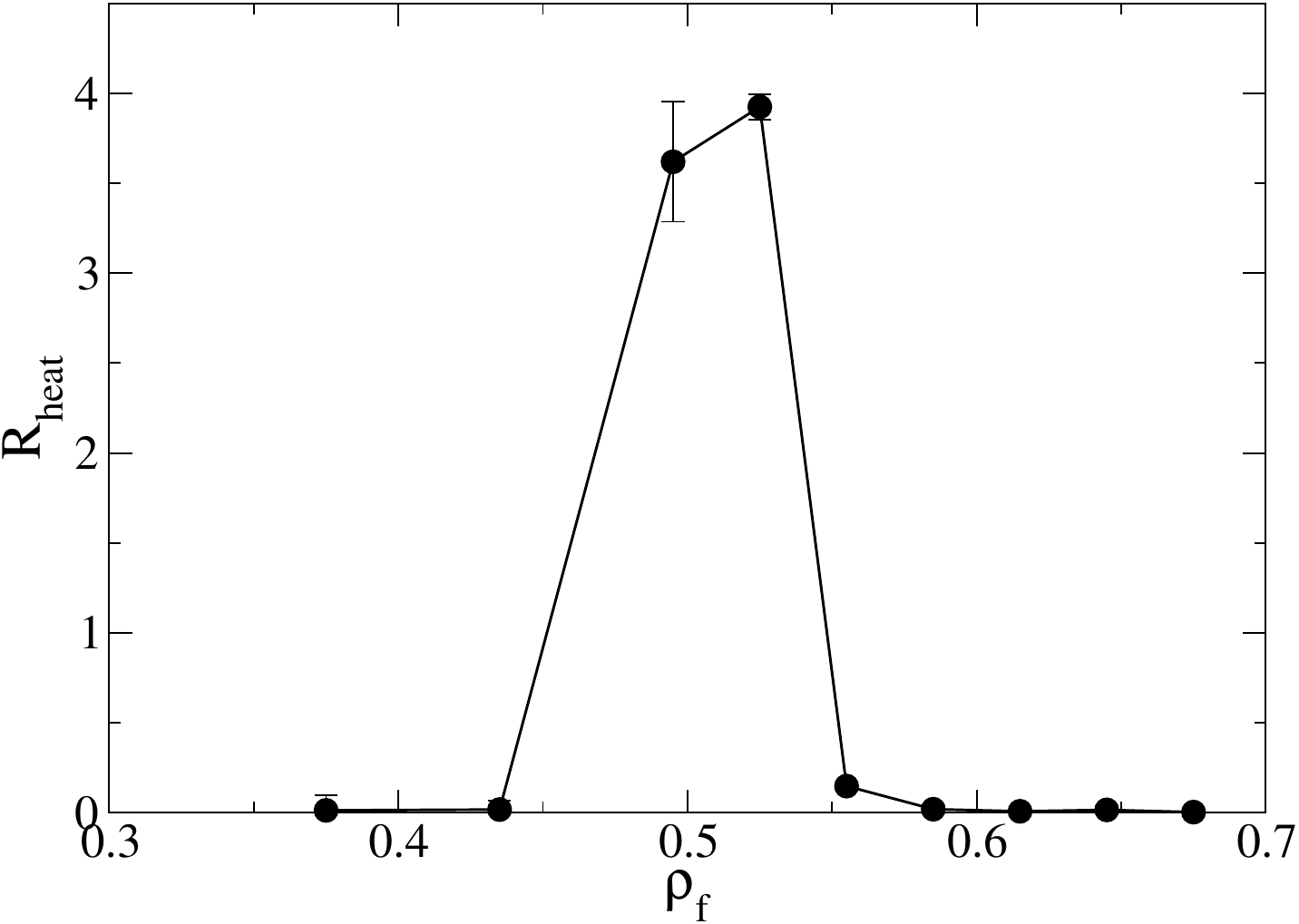}

\caption{Rectification coefficient $R_{ \rm heat}$ as a function of channel filling $\rho_{f}$ for flexible polymers. We find a range of fluid fillings which, in combination with the polymers result in a significative asymmetry
in the chamber heat transfer when hot and cold walls are swapped in
the channel. \protect\label{fig:Rectification-coefficient-flexible_polymers}}
\end{figure}

In Fig. \ref{fig:Rectification-coefficient-flexible_polymers}, the rectification coefficient $R_{\rm heat}$ is plotted as a function of $\rho_{f}$. A significant rectification is clearly observed in the filling densities range $\rho_{f}\in[0.44\sigma^{-3},\,0.58\sigma^{-3}]$.
In this case, the heat flowing in DM is up to five times higher than that of IM.
For low and high fillings the diode effect vanishes and the  interval of rectification is narrower than that of  stiff polymers. 

\begin{figure}
\includegraphics[width=0.80\columnwidth]{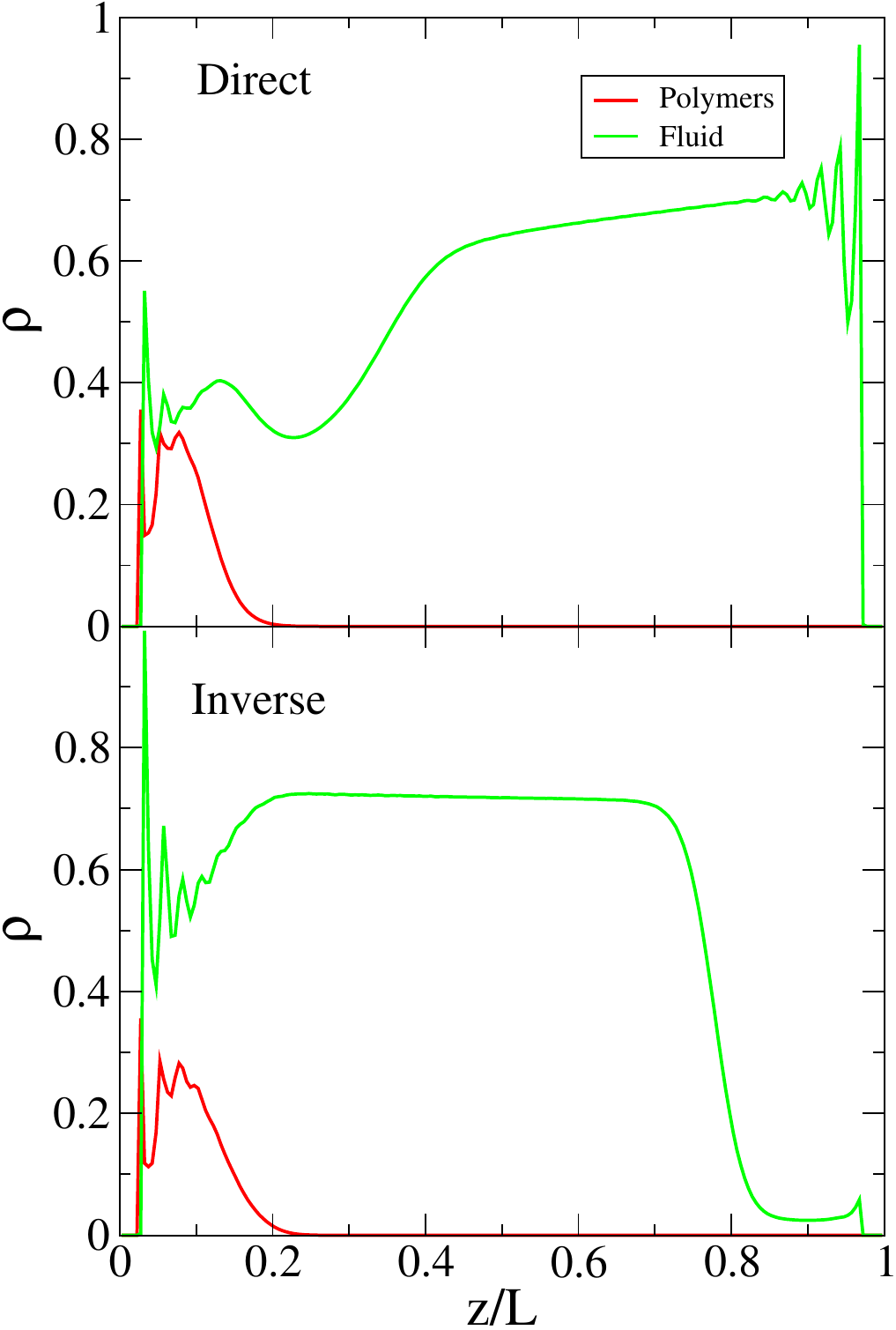}

\caption{Density profile of the channel in DM (upper panel) and IM (lower panel) for a fluid filling of $\rho_{f}=0.453\sigma^{-3}$. The fluid is indicated in green lines  and the density of grafted polymers is shown in red lines.  \protect\label{fig:Density-flex_middle_filling}}
\end{figure}

For low filling   density, we observe a vapor phase close to the hot wall in both DM and IM. These cases are presented in Supplementary Material.  In Fig. S1 we present the density profiles for a density  of $\rho_f=0.435\sigma^{-3}$ (low filling density). For high filling densities there is no vapor phase and a liquid is observed all-across the chamber, with decreasing density towards the hot wall, following the increase of temperature. In  Fig. S2, we present the density profiles for a high-density case ($\rho_f=0.585\sigma^{-3}$) for flexible polymers. Both cases present vanishing rectification coefficients
We present here  the density prof iles for intermediate filling density ($\rho_{f}=0.453\sigma^{-3}$), corresponding to the peak of rectification for a nano-chamber coated with flexible polymers.  Figure \ref{fig:Density-flex_middle_filling} has a significant asymmetry when the diode is set in DM (upper panel) or IM (lower panel). 
The IM has a similar structure and is also observed for the lower density case (see Fig. S1 in SM). There is a clear low-density vapor phase, and the liquid wets the cold wall, together with the grafted polymers. The DM (upper panel, Fig. \ref{fig:Density-flex_middle_filling}) presents a very different structure. The polymer interacts with the liquid and interfacial zones, leading to a significantly more homogeneous density distribution across the chamber. There are no low-density zones, and therefore, no region of high thermal resistance. The visual inspection of the simulations (see as an example the right panel of Fig. \ref{fig:system_sample}) shows that the middle density region, in the interval $z/L\in[0.75,1.0]$ corresponds to a combination of two fluid morphologies. On one hand there are  liquid-like zones, which wet the polymers and the hot wall. And these liquid structures coexist with   vapor bubbles. The presence of liquid-like structures is responsible for a much higher thermal conductivity of the overall chamber, much higher than that of the IM. This filling has a high asymmetry when comparing DM (low thermal resistance) and IM (high thermal resistance). The combination of this fluid filling and the grafted flexible polymers allows for significant heat rectification. 

The density profiles in Figure \ref{fig:Density-flex_middle_filling} and those of Fig. S1
and S2 shown in SM,  allow us to find the key elements for heat rectification in the case of flexible polymers.
On one hand, the fluid density must be low enough to ensure the presence of a liquid-vapor interface when the cold wall is set in the polymer-coated wall (IM).  On the other hand, when the channel is set in the DM, the filling density has to be high enough such that the liquid-vapor interface is relatively close to the polymer layer. If this happens, from the interaction among the polymers with the fluid particles a partial wetting of the hot wall occurs, which gives rise to regions of liquid reaching the hot wall coexisting with vapor bubbles.
This results in a thermal resistance significantly lower than that of the pure vapor region, which is present in inverse mode.

\begin{figure}
\includegraphics[width=0.98\columnwidth]{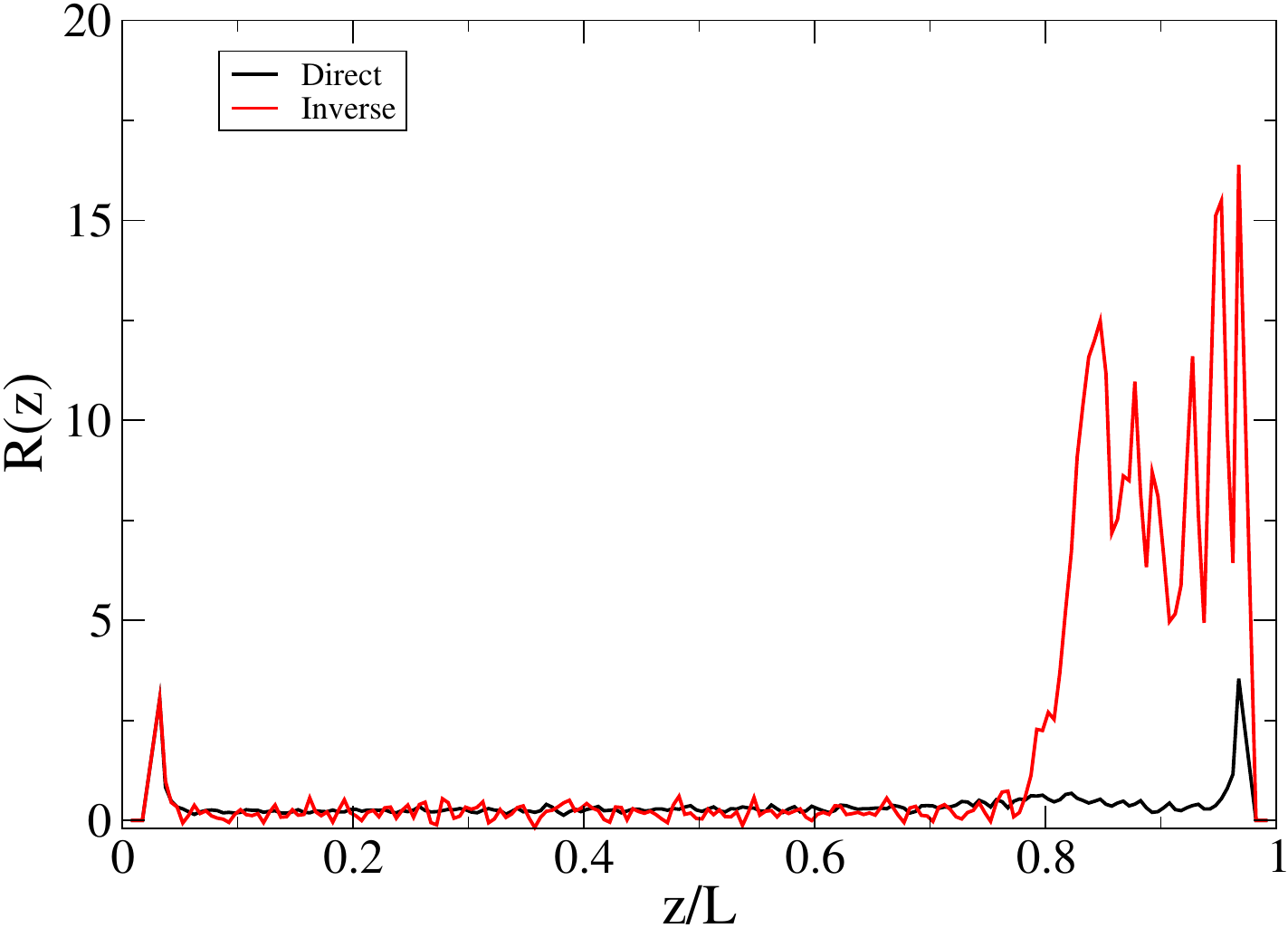}

\caption{Resistivity profile  $\rho_{f}=0.526\sigma^{-3}$
 which presents the maximum rectification factor for
flexible polymers. The profiles were adjusted to have the cold wall
at $z/L=0$ and the hot wall at $z/L=1$. The direct mode
is shown in black line and the inverse mode in red line. The low thermal resistance
zone corresponds to the liquid phase and the highly fluctuating part,
to the vapor region of the chamber. \protect\label{fig:Resistance-flexible_polymers}}
\end{figure}

In Figure \ref{fig:Resistance-flexible_polymers}, we show the resistivity profiles in DM (black lines)
and IM (red line) modes for a filling density of $\rho_{f}=0.526\sigma^{-3}$, which corresponds
to the maximum rectification factor for flexible polymers. The resistivity profile for a case of low filling density,
which presents vanishing rectification coefficient is presented and explained in Fig. S5 (SM). 
One of the profiles is inverted in positions, such that the cold wall is always located in $z/L=0$ (in contact with the liquid) and the hot wall is located at $z/L=1$. The liquid phase presents, in both modes,
a similar value and very low thermal resistance. 
 After that, a peak is observed, only for IM (red line), close to the location of the liquid-vapor interface. These
peak corresponds the Kapitza resistance\citep{Pastorino2022}, already explained for stiff polymers.
As discussed for stiff polymers, the vapor zone is characterized by a noisy signal of high resistance. The higher fluctuations are due to the poorer statistics, because the temperature is averaged over much fewer particles than that of the liquid, in this low-density vapor zone. 
This region has resistance much higher than that of the liquid and dominates
the overall thermal behavior of the nano-chamber.  There is a different polymer-fluid structure when the chamber is set either in DM or IM. In IM, it is mainly a vapor region with very high resistance and another peak, close to
the wall, which is a vapor-wall Kapitza resistance (Fig. \ref{fig:Resistance-flexible_polymers},
red line, upper panel). 
The main feature is a huge difference in the vapor zone. In DM the resistance keeps on being very low and similar
to that of the liquid phase. Two small peaks are present which, as in the case of stiff polymers, 
are the Kapitza resistance between wall and fluid in both extremes of the nano-chamber. 
Here the Kapitza resistance for the liquid-vapor interface
is practically washed out and the whole chamber has a low value liquid-like
resistance. The coexistence of vapor bubbles with liquid-like zones,
favored by the polymers, is responsible for this
low resistance. See, for example, the right panel of Fig. \ref{fig:system_sample}. 
The overall channel resistance is therefore much higher in IM
than in DM, giving rise to a high rectification factor (see
Fig. \ref{fig:Rectification-coefficient-flexible_polymers}) for flexible polymers. 

\section{Final discussion and conclusions\protect\label{sec:Conclusions}}

In this work, we propose a thermal diode based on a nano-chamber coated with end-grafted polymers and partially filled with a fluid in liquid-vapor coexistence. We found that when a temperature gradient is applied to the system with a low density of end-grafted polymers, the device can exhibit a high degree of heat rectification, provided it is filled with an appropriate amount of fluid in a liquid–vapor coexistence regime.
According to recent studies thermal diodes designed with two plates and hydrophobic/hydrophilic coatings are being used to confine a fluid in phase coexistence. 
For temperature ranges similar to ours, these diodes present rectification coefficients of around  $\lesssim 10 $\cite{ZHAO23, Avanessian2016}.  Our proposal, which is based on a different mechanism, involving coating polymers onto one of the plates,  presents rectification coefficients with similar values 
for both highly rigid and fully flexible polymers.

When the polymers are grafted onto the hot wall they induce partial wetting.
This interaction leads to the disruption of the otherwise homogeneous vapor
layer across a region of the nano-chamber.  This mechanism is present,
independently of the polymer morphology.  In direct mode, polymers form liquid
threads of high liquid-like thermal conductivity which coexist with vapor
bubbles. In inverse mode, the vapor layer is present across the entire
nano-chamber. This creates a zone of very high thermal resistivity, leading to
a low heat flux.   
effect occurs is dependent on several factors, including polymer stiffness,
grafting density, polymer chain length and fluid-polymer interaction, which are
essentially nanoscopic in nature. 
It is important to note that the rectification effect studied here, while
tested in a nano-chamber, is very versatile and scalable towards macroscopic
length scales in the lateral direction, i.e. perpendicular to the thermal
gradient. The fluid filling range for which the thermal diode effect  takes
place depends on several factors such as: polymer stiffness, grafting density,
polymers chain length and fluid-polymer interaction, which are essentially
nanoscopic. The vapor phase in inverse mode should have the length scale of the
polymer layer and therefore,  it should be  thinner for more flexible polymers.
The intermediate values of filling density maximize the rectification factor,
but thermal rectification is also found at low filling density for stiff
polymers. However, the overall heat resistivity is very high and the heat flux
are quite small.  We also  note that, while we found a twice higher
rectification factor for stiff  than for fully flexible polymers, the latter
case is very important because it extends the range of applications of the
thermal diode effect for practically  any  choice of polymeric coating.  We
expect that the results we found here can motivate further experimental and
theoretical works. 

Our proposal for thermal rectifiers based on grafted polymer coatings is  highly versatile and it has the potential to be used as a device in a variety of thermal architectures at different scales and when high diodicity is required.

\begin{acknowledgments}
C. P.  thanks CONICET for partial support of this work through grant PIP 11220210100546CO. CNEA and SECYT are also gratefully acknowledged for providing  supercomputing resources through for our LabSim Supercomputing Lab. 
M. F .C and A. M thanks to CyTUNGS 30/1161 and to PIP-CONICET 11220200101599CO]
\end{acknowledgments}


\begin{thebibliography}{51}%
\makeatletter
\providecommand \@ifxundefined [1]{%
 \@ifx{#1\undefined}
}%
\providecommand \@ifnum [1]{%
 \ifnum #1\expandafter \@firstoftwo
 \else \expandafter \@secondoftwo
 \fi
}%
\providecommand \@ifx [1]{%
 \ifx #1\expandafter \@firstoftwo
 \else \expandafter \@secondoftwo
 \fi
}%
\providecommand \natexlab [1]{#1}%
\providecommand \enquote  [1]{``#1''}%
\providecommand \bibnamefont  [1]{#1}%
\providecommand \bibfnamefont [1]{#1}%
\providecommand \citenamefont [1]{#1}%
\providecommand \href@noop [0]{\@secondoftwo}%
\providecommand \href [0]{\begingroup \@sanitize@url \@href}%
\providecommand \@href[1]{\@@startlink{#1}\@@href}%
\providecommand \@@href[1]{\endgroup#1\@@endlink}%
\providecommand \@sanitize@url [0]{\catcode `\\12\catcode `\$12\catcode
  `\&12\catcode `\#12\catcode `\^12\catcode `\_12\catcode `\%12\relax}%
\providecommand \@@startlink[1]{}%
\providecommand \@@endlink[0]{}%
\providecommand \url  [0]{\begingroup\@sanitize@url \@url }%
\providecommand \@url [1]{\endgroup\@href {#1}{\urlprefix }}%
\providecommand \urlprefix  [0]{URL }%
\providecommand \Eprint [0]{\href }%
\providecommand \doibase [0]{http://dx.doi.org/}%
\providecommand \selectlanguage [0]{\@gobble}%
\providecommand \bibinfo  [0]{\@secondoftwo}%
\providecommand \bibfield  [0]{\@secondoftwo}%
\providecommand \translation [1]{[#1]}%
\providecommand \BibitemOpen [0]{}%
\providecommand \bibitemStop [0]{}%
\providecommand \bibitemNoStop [0]{.\EOS\space}%
\providecommand \EOS [0]{\spacefactor3000\relax}%
\providecommand \BibitemShut  [1]{\csname bibitem#1\endcsname}%
\let\auto@bib@innerbib\@empty
\bibitem [{\citenamefont {Hassan}\ \emph {et~al.}(2018)\citenamefont {Hassan},
  \citenamefont {Savaria},\ and\ \citenamefont {Sawan}}]{Hassan18}%
  \BibitemOpen
  \bibfield  {author} {\bibinfo {author} {\bibfnamefont {A.}~\bibnamefont
  {Hassan}}, \bibinfo {author} {\bibfnamefont {Y.}~\bibnamefont {Savaria}}, \
  and\ \bibinfo {author} {\bibfnamefont {M.}~\bibnamefont {Sawan}},\ }\href
  {\doibase 10.1109/TVLSI.2018.2834499} {\bibfield  {journal} {\bibinfo
  {journal} {IEEE Transactions on Very Large Scale Integration (VLSI) Systems}\
  }\textbf {\bibinfo {volume} {26}},\ \bibinfo {pages} {2085} (\bibinfo {year}
  {2018})}\BibitemShut {NoStop}%
\bibitem [{\citenamefont {Dhumal}\ \emph {et~al.}(2023)\citenamefont {Dhumal},
  \citenamefont {Kulkarni},\ and\ \citenamefont {Ambhore}}]{Dhumal23}%
  \BibitemOpen
  \bibfield  {author} {\bibinfo {author} {\bibfnamefont {R.}~\bibnamefont
  {Dhumal}, \bibfnamefont {Amol}}, \bibinfo {author} {\bibfnamefont
  {P.}~\bibnamefont {Kulkarni}, \bibfnamefont {Atul}}, \ and\ \bibinfo {author}
  {\bibfnamefont {N.~H.}\ \bibnamefont {Ambhore}},\ }\href {\doibase
  10.1186/s44147-023-00309-2} {\bibfield  {journal} {\bibinfo  {journal}
  {Journal of Engineering and Applied Science}\ }\textbf {\bibinfo {volume}
  {140,70}},\ \bibinfo {pages} {2536} (\bibinfo {year} {2023})}\BibitemShut
  {NoStop}%
\bibitem [{\citenamefont {Smoyer}\ and\ \citenamefont {and}(2019)}]{Smoyer19}%
  \BibitemOpen
  \bibfield  {author} {\bibinfo {author} {\bibfnamefont {J.~L.}\ \bibnamefont
  {Smoyer}}\ and\ \bibinfo {author} {\bibfnamefont {P.~M.~N.}\ \bibnamefont
  {and}},\ }\href {\doibase 10.1080/01457632.2018.1426265} {\bibfield
  {journal} {\bibinfo  {journal} {Heat Transfer Engineering}\ }\textbf
  {\bibinfo {volume} {40}},\ \bibinfo {pages} {269} (\bibinfo {year}
  {2019})}\BibitemShut {NoStop}%
\bibitem [{\citenamefont {Y.}\ \emph {et~al.}(2023)\citenamefont {Y.},
  \citenamefont {M.}, \citenamefont {T.}, \citenamefont {J.}, \citenamefont
  {J.},\ and\ \citenamefont {H.}}]{Jung23}%
  \BibitemOpen
  \bibfield  {author} {\bibinfo {author} {\bibfnamefont {J.}~\bibnamefont
  {Y.}}, \bibinfo {author} {\bibfnamefont {K.}~\bibnamefont {M.}}, \bibinfo
  {author} {\bibfnamefont {K.}~\bibnamefont {T.}}, \bibinfo {author}
  {\bibfnamefont {A.}~\bibnamefont {J.}}, \bibinfo {author} {\bibfnamefont
  {L.}~\bibnamefont {J.}}, \ and\ \bibinfo {author} {\bibfnamefont {K.~S.}\
  \bibnamefont {H.}},\ }\href {\doibase 10.1007/s40820-023-01126-1} {\bibfield
  {journal} {\bibinfo  {journal} {Nano-Micro Letters}\ }\textbf {\bibinfo
  {volume} {160,15}} (\bibinfo {year} {2023}),\
  10.1007/s40820-023-01126-1}\BibitemShut {NoStop}%
\bibitem [{\citenamefont {Jayathilaka}\ \emph {et~al.}(2019)\citenamefont
  {Jayathilaka}, \citenamefont {Qi}, \citenamefont {Qin}, \citenamefont
  {Chinnappan}, \citenamefont {Serrano-García}, \citenamefont {Baskar},
  \citenamefont {Wang}, \citenamefont {He}, \citenamefont {Cui}, \citenamefont
  {Thomas},\ and\ \citenamefont {Ramakrishna}}]{Jaya19}%
  \BibitemOpen
  \bibfield  {author} {\bibinfo {author} {\bibfnamefont {W.~A. D.~M.}\
  \bibnamefont {Jayathilaka}}, \bibinfo {author} {\bibfnamefont
  {K.}~\bibnamefont {Qi}}, \bibinfo {author} {\bibfnamefont {Y.}~\bibnamefont
  {Qin}}, \bibinfo {author} {\bibfnamefont {A.}~\bibnamefont {Chinnappan}},
  \bibinfo {author} {\bibfnamefont {W.}~\bibnamefont {Serrano-García}},
  \bibinfo {author} {\bibfnamefont {C.}~\bibnamefont {Baskar}}, \bibinfo
  {author} {\bibfnamefont {H.}~\bibnamefont {Wang}}, \bibinfo {author}
  {\bibfnamefont {J.}~\bibnamefont {He}}, \bibinfo {author} {\bibfnamefont
  {S.}~\bibnamefont {Cui}}, \bibinfo {author} {\bibfnamefont {S.~W.}\
  \bibnamefont {Thomas}}, \ and\ \bibinfo {author} {\bibfnamefont
  {S.}~\bibnamefont {Ramakrishna}},\ }\href {\doibase
  https://doi.org/10.1002/adma.201805921} {\bibfield  {journal} {\bibinfo
  {journal} {Advanced Materials}\ }\textbf {\bibinfo {volume} {31}},\ \bibinfo
  {pages} {1805921} (\bibinfo {year} {2019})}\BibitemShut {NoStop}%
\bibitem [{\citenamefont {Bar-Cohen}\ and\ \citenamefont
  {Wang}(2021)}]{BarCohen21}%
  \BibitemOpen
  \bibfield  {author} {\bibinfo {author} {\bibfnamefont {A.}~\bibnamefont
  {Bar-Cohen}}\ and\ \bibinfo {author} {\bibfnamefont {P.}~\bibnamefont
  {Wang}},\ }\enquote {\bibinfo {title} {On-chip thermal management and
  hot-spot remediation},}\ in\ \href {\doibase 10.1007/978-3-030-49991-4_9}
  {\emph {\bibinfo {booktitle} {Nano-Bio- Electronic, Photonic and MEMS
  Packaging}}},\ \bibinfo {editor} {edited by\ \bibinfo {editor} {\bibfnamefont
  {C.~P.-P.}\ \bibnamefont {Wong}}, \bibinfo {editor} {\bibfnamefont
  {K.-s.~J.}\ \bibnamefont {Moon}}, \ and\ \bibinfo {editor} {\bibfnamefont
  {Y.}~\bibnamefont {Li}}}\ (\bibinfo  {publisher} {Springer International
  Publishing},\ \bibinfo {address} {Cham},\ \bibinfo {year} {2021})\ pp.\
  \bibinfo {pages} {157--203}\BibitemShut {NoStop}%
\bibitem [{\citenamefont {Tachikawa}\ \emph {et~al.}(2022)\citenamefont
  {Tachikawa}, \citenamefont {Nagano}, \citenamefont {Ohnishi},\ and\
  \citenamefont {Nagasaka}}]{Tachikawa22}%
  \BibitemOpen
  \bibfield  {author} {\bibinfo {author} {\bibfnamefont {S.}~\bibnamefont
  {Tachikawa}}, \bibinfo {author} {\bibfnamefont {H.}~\bibnamefont {Nagano}},
  \bibinfo {author} {\bibfnamefont {A.}~\bibnamefont {Ohnishi}}, \ and\
  \bibinfo {author} {\bibfnamefont {Y.}~\bibnamefont {Nagasaka}},\ }\href
  {\doibase 10.1007/s10765-022-03010-3} {\bibfield  {journal} {\bibinfo
  {journal} {Int J Thermophys}\ }\textbf {\bibinfo {volume} {43}} (\bibinfo
  {year} {2022}),\ 10.1007/s10765-022-03010-3}\BibitemShut {NoStop}%
\bibitem [{\citenamefont {Wehmeyer}\ \emph {et~al.}(2017)\citenamefont
  {Wehmeyer}, \citenamefont {Yabuki}, \citenamefont {Monachon}, \citenamefont
  {Wu},\ and\ \citenamefont {Dames}}]{Wehmeyer17}%
  \BibitemOpen
  \bibfield  {author} {\bibinfo {author} {\bibfnamefont {G.}~\bibnamefont
  {Wehmeyer}}, \bibinfo {author} {\bibfnamefont {T.}~\bibnamefont {Yabuki}},
  \bibinfo {author} {\bibfnamefont {C.}~\bibnamefont {Monachon}}, \bibinfo
  {author} {\bibfnamefont {J.}~\bibnamefont {Wu}}, \ and\ \bibinfo {author}
  {\bibfnamefont {C.}~\bibnamefont {Dames}},\ }\href {\doibase
  10.1063/1.5001072} {\bibfield  {journal} {\bibinfo  {journal} {Applied
  Physics Reviews}\ }\textbf {\bibinfo {volume} {4}},\ \bibinfo {pages}
  {041304} (\bibinfo {year} {2017})},\ \Eprint
  {http://arxiv.org/abs/https://pubs.aip.org/aip/apr/article-pdf/doi/10.1063/1.5001072/14574738/041304\_1\_online.pdf}
  {https://pubs.aip.org/aip/apr/article-pdf/doi/10.1063/1.5001072/14574738/041304\_1\_online.pdf}
  \BibitemShut {NoStop}%
\bibitem [{\citenamefont {Cho}\ and\ \citenamefont {Gabbar}(2019)}]{Cho19}%
  \BibitemOpen
  \bibfield  {author} {\bibinfo {author} {\bibfnamefont {Y.}~\bibnamefont
  {Cho}}\ and\ \bibinfo {author} {\bibfnamefont {H.~A.}\ \bibnamefont
  {Gabbar}},\ }\href {\doibase 10.1007/s42797-019-00002-9} {\bibfield
  {journal} {\bibinfo  {journal} {Safety in Extreme Environments}\ ,\ \bibinfo
  {pages} {11}} (\bibinfo {year} {2019})}\BibitemShut {NoStop}%
\bibitem [{\citenamefont {Pang}\ \emph {et~al.}(2024)\citenamefont {Pang},
  \citenamefont {Li}, \citenamefont {Wen}, \citenamefont {Liang}, \citenamefont
  {Gao}, \citenamefont {Yang}, \citenamefont {Huang}, \citenamefont {Xu},
  \citenamefont {Luo}, \citenamefont {Zeng},\ and\ \citenamefont
  {Sun}}]{Pang24}%
  \BibitemOpen
  \bibfield  {author} {\bibinfo {author} {\bibfnamefont {Y.}~\bibnamefont
  {Pang}}, \bibinfo {author} {\bibfnamefont {J.}~\bibnamefont {Li}}, \bibinfo
  {author} {\bibfnamefont {Z.}~\bibnamefont {Wen}}, \bibinfo {author}
  {\bibfnamefont {T.}~\bibnamefont {Liang}}, \bibinfo {author} {\bibfnamefont
  {S.}~\bibnamefont {Gao}}, \bibinfo {author} {\bibfnamefont {M.}~\bibnamefont
  {Yang}}, \bibinfo {author} {\bibfnamefont {D.}~\bibnamefont {Huang}},
  \bibinfo {author} {\bibfnamefont {J.}~\bibnamefont {Xu}}, \bibinfo {author}
  {\bibfnamefont {T.}~\bibnamefont {Luo}}, \bibinfo {author} {\bibfnamefont
  {X.}~\bibnamefont {Zeng}}, \ and\ \bibinfo {author} {\bibfnamefont
  {R.}~\bibnamefont {Sun}},\ }\href {\doibase
  https://doi.org/10.1016/j.mtphys.2024.101450} {\bibfield  {journal} {\bibinfo
   {journal} {Materials Today Physics}\ }\textbf {\bibinfo {volume} {44}},\
  \bibinfo {pages} {101450} (\bibinfo {year} {2024})}\BibitemShut {NoStop}%
\bibitem [{\citenamefont {Bianco}\ \emph {et~al.}(2022)\citenamefont {Bianco},
  \citenamefont {{De Rosa}},\ and\ \citenamefont {Vafai}}]{Bianco22}%
  \BibitemOpen
  \bibfield  {author} {\bibinfo {author} {\bibfnamefont {V.}~\bibnamefont
  {Bianco}}, \bibinfo {author} {\bibfnamefont {M.}~\bibnamefont {{De Rosa}}}, \
  and\ \bibinfo {author} {\bibfnamefont {K.}~\bibnamefont {Vafai}},\ }\href
  {\doibase https://doi.org/10.1016/j.applthermaleng.2022.118839} {\bibfield
  {journal} {\bibinfo  {journal} {Applied Thermal Engineering}\ }\textbf
  {\bibinfo {volume} {214}},\ \bibinfo {pages} {118839} (\bibinfo {year}
  {2022})}\BibitemShut {NoStop}%
\bibitem [{\citenamefont {Jun}\ \emph {et~al.}(2025)\citenamefont {Jun},
  \citenamefont {Haiyang}, \citenamefont {Guodong}, \citenamefont {Xiaoping},\
  and\ \citenamefont {Xiangjun}}]{JUN2025}%
  \BibitemOpen
  \bibfield  {author} {\bibinfo {author} {\bibfnamefont {W.}~\bibnamefont
  {Jun}}, \bibinfo {author} {\bibfnamefont {L.}~\bibnamefont {Haiyang}},
  \bibinfo {author} {\bibfnamefont {X.}~\bibnamefont {Guodong}}, \bibinfo
  {author} {\bibfnamefont {W.}~\bibnamefont {Xiaoping}}, \ and\ \bibinfo
  {author} {\bibfnamefont {C.}~\bibnamefont {Xiangjun}},\ }\href {\doibase
  https://doi.org/10.1016/j.icheatmasstransfer.2024.108517} {\bibfield
  {journal} {\bibinfo  {journal} {International Communications in Heat and Mass
  Transfer}\ }\textbf {\bibinfo {volume} {161}},\ \bibinfo {pages} {108517}
  (\bibinfo {year} {2025})}\BibitemShut {NoStop}%
\bibitem [{\citenamefont {Swoboda}\ \emph {et~al.}(2021)\citenamefont
  {Swoboda}, \citenamefont {Klinar}, \citenamefont {Yalamarthy}, \citenamefont
  {Kitanovski},\ and\ \citenamefont {Mu\~noz Rojo}}]{Swoboda21}%
  \BibitemOpen
  \bibfield  {author} {\bibinfo {author} {\bibfnamefont {T.}~\bibnamefont
  {Swoboda}}, \bibinfo {author} {\bibfnamefont {K.}~\bibnamefont {Klinar}},
  \bibinfo {author} {\bibfnamefont {A.~S.}\ \bibnamefont {Yalamarthy}},
  \bibinfo {author} {\bibfnamefont {A.}~\bibnamefont {Kitanovski}}, \ and\
  \bibinfo {author} {\bibfnamefont {M.}~\bibnamefont {Mu\~noz Rojo}},\ }\href
  {\doibase 10.1002/aelm.202000625} {\bibfield  {journal} {\bibinfo  {journal}
  {Advanced Electronic Materials}\ }\textbf {\bibinfo {volume} {7}},\ \bibinfo
  {pages} {2000625} (\bibinfo {year} {2021})}\BibitemShut {NoStop}%
\bibitem [{\citenamefont {Wong}\ \emph {et~al.}(2021)\citenamefont {Wong},
  \citenamefont {Tso}, \citenamefont {Ho},\ and\ \citenamefont {Lee}}]{Wong21}%
  \BibitemOpen
  \bibfield  {author} {\bibinfo {author} {\bibfnamefont {M.}~\bibnamefont
  {Wong}}, \bibinfo {author} {\bibfnamefont {C.}~\bibnamefont {Tso}}, \bibinfo
  {author} {\bibfnamefont {T.}~\bibnamefont {Ho}}, \ and\ \bibinfo {author}
  {\bibfnamefont {H.}~\bibnamefont {Lee}},\ }\href {\doibase
  /10.1016/j.ijheatmasstransfer.2020.120607} {\bibfield  {journal} {\bibinfo
  {journal} {International Journal of Heat and Mass Transfer}\ }\textbf
  {\bibinfo {volume} {164}},\ \bibinfo {pages} {120607} (\bibinfo {year}
  {2021})}\BibitemShut {NoStop}%
\bibitem [{\citenamefont {Malik}\ and\ \citenamefont
  {Fobelets}(2022)}]{malik22}%
  \BibitemOpen
  \bibfield  {author} {\bibinfo {author} {\bibfnamefont {F.~K.}\ \bibnamefont
  {Malik}}\ and\ \bibinfo {author} {\bibfnamefont {K.}~\bibnamefont
  {Fobelets}},\ }\href@noop {} {\bibfield  {journal} {\bibinfo  {journal}
  {Journal of Semiconductors}\ }\textbf {\bibinfo {volume} {43}},\ \bibinfo
  {pages} {103101} (\bibinfo {year} {2022})}\BibitemShut {NoStop}%
\bibitem [{\citenamefont {Zurdo}\ \emph
  {et~al.}(2024{\natexlab{a}})\citenamefont {Zurdo}, \citenamefont {Chej},
  \citenamefont {Monastra},\ and\ \citenamefont {Carusela}}]{zurdo24}%
  \BibitemOpen
  \bibfield  {author} {\bibinfo {author} {\bibfnamefont {L.~L.}\ \bibnamefont
  {Zurdo}}, \bibinfo {author} {\bibfnamefont {L.~G.}\ \bibnamefont {Chej}},
  \bibinfo {author} {\bibfnamefont {A.~G.}\ \bibnamefont {Monastra}}, \ and\
  \bibinfo {author} {\bibfnamefont {M.~F.}\ \bibnamefont {Carusela}},\ }\href
  {\doibase https://doi.org/10.1016/j.ijheatmasstransfer.2023.125110}
  {\bibfield  {journal} {\bibinfo  {journal} {International Journal of Heat and
  Mass Transfer}\ }\textbf {\bibinfo {volume} {222}},\ \bibinfo {pages}
  {125110} (\bibinfo {year} {2024}{\natexlab{a}})}\BibitemShut {NoStop}%
\bibitem [{\citenamefont {Wang}\ \emph {et~al.}(2017)\citenamefont {Wang},
  \citenamefont {Hu}, \citenamefont {Takahashi}, \citenamefont {Zhang},
  \citenamefont {Takamatsu},\ and\ \citenamefont {Chen}}]{wang2017}%
  \BibitemOpen
  \bibfield  {author} {\bibinfo {author} {\bibfnamefont {H.}~\bibnamefont
  {Wang}}, \bibinfo {author} {\bibfnamefont {S.}~\bibnamefont {Hu}}, \bibinfo
  {author} {\bibfnamefont {K.}~\bibnamefont {Takahashi}}, \bibinfo {author}
  {\bibfnamefont {X.}~\bibnamefont {Zhang}}, \bibinfo {author} {\bibfnamefont
  {H.}~\bibnamefont {Takamatsu}}, \ and\ \bibinfo {author} {\bibfnamefont
  {J.}~\bibnamefont {Chen}},\ }\href@noop {} {\bibfield  {journal} {\bibinfo
  {journal} {Nature communications}\ }\textbf {\bibinfo {volume} {8}},\
  \bibinfo {pages} {15843} (\bibinfo {year} {2017})}\BibitemShut {NoStop}%
\bibitem [{\citenamefont {Yousefi}\ \emph {et~al.}(2020)\citenamefont
  {Yousefi}, \citenamefont {Khoeini},\ and\ \citenamefont
  {Rajabpour}}]{yousefi2020}%
  \BibitemOpen
  \bibfield  {author} {\bibinfo {author} {\bibfnamefont {F.}~\bibnamefont
  {Yousefi}}, \bibinfo {author} {\bibfnamefont {F.}~\bibnamefont {Khoeini}}, \
  and\ \bibinfo {author} {\bibfnamefont {A.}~\bibnamefont {Rajabpour}},\
  }\href@noop {} {\bibfield  {journal} {\bibinfo  {journal} {International
  Journal of Heat and Mass Transfer}\ }\textbf {\bibinfo {volume} {146}},\
  \bibinfo {pages} {118884} (\bibinfo {year} {2020})}\BibitemShut {NoStop}%
\bibitem [{\citenamefont {Carlomagno}\ \emph {et~al.}(2020)\citenamefont
  {Carlomagno}, \citenamefont {Cimmelli},\ and\ \citenamefont
  {Jou}}]{carlomagno2020}%
  \BibitemOpen
  \bibfield  {author} {\bibinfo {author} {\bibfnamefont {I.}~\bibnamefont
  {Carlomagno}}, \bibinfo {author} {\bibfnamefont {V.}~\bibnamefont
  {Cimmelli}}, \ and\ \bibinfo {author} {\bibfnamefont {D.}~\bibnamefont
  {Jou}},\ }\href@noop {} {\bibfield  {journal} {\bibinfo  {journal} {Mechanics
  Research Communications}\ }\textbf {\bibinfo {volume} {103}},\ \bibinfo
  {pages} {103472} (\bibinfo {year} {2020})}\BibitemShut {NoStop}%
\bibitem [{\citenamefont {Liu}\ \emph {et~al.}(2019)\citenamefont {Liu},
  \citenamefont {Wang},\ and\ \citenamefont {Zhang}}]{Liu2019}%
  \BibitemOpen
  \bibfield  {author} {\bibinfo {author} {\bibfnamefont {H.}~\bibnamefont
  {Liu}}, \bibinfo {author} {\bibfnamefont {H.}~\bibnamefont {Wang}}, \ and\
  \bibinfo {author} {\bibfnamefont {X.}~\bibnamefont {Zhang}},\ }\href
  {\doibase 10.3390/app9020344} {\bibfield  {journal} {\bibinfo  {journal}
  {Applied Sciences}\ }\textbf {\bibinfo {volume} {9}} (\bibinfo {year}
  {2019}),\ 10.3390/app9020344}\BibitemShut {NoStop}%
\bibitem [{\citenamefont {Zeng}\ and\ \citenamefont {Wang}(2008)}]{zeng2008}%
  \BibitemOpen
  \bibfield  {author} {\bibinfo {author} {\bibfnamefont {N.}~\bibnamefont
  {Zeng}}\ and\ \bibinfo {author} {\bibfnamefont {J.-S.}\ \bibnamefont
  {Wang}},\ }\href@noop {} {\bibfield  {journal} {\bibinfo  {journal} {Physical
  Review B—Condensed Matter and Materials Physics}\ }\textbf {\bibinfo
  {volume} {78}},\ \bibinfo {pages} {024305} (\bibinfo {year}
  {2008})}\BibitemShut {NoStop}%
\bibitem [{\citenamefont {Zurdo}\ \emph
  {et~al.}(2024{\natexlab{b}})\citenamefont {Zurdo}, \citenamefont {Chej},
  \citenamefont {Monastra},\ and\ \citenamefont {Carusela}}]{zurdo2024}%
  \BibitemOpen
  \bibfield  {author} {\bibinfo {author} {\bibfnamefont {L.~L.}\ \bibnamefont
  {Zurdo}}, \bibinfo {author} {\bibfnamefont {L.~G.}\ \bibnamefont {Chej}},
  \bibinfo {author} {\bibfnamefont {A.~G.}\ \bibnamefont {Monastra}}, \ and\
  \bibinfo {author} {\bibfnamefont {M.~F.}\ \bibnamefont {Carusela}},\
  }\href@noop {} {\bibfield  {journal} {\bibinfo  {journal} {International
  Journal of Heat and Mass Transfer}\ }\textbf {\bibinfo {volume} {222}},\
  \bibinfo {pages} {125110} (\bibinfo {year} {2024}{\natexlab{b}})}\BibitemShut
  {NoStop}%
\bibitem [{\citenamefont {L{\'o}pez-Su{\'a}rez}\ \emph
  {et~al.}(2018)\citenamefont {L{\'o}pez-Su{\'a}rez}, \citenamefont {Neri},\
  and\ \citenamefont {Rurali}}]{lopez2018}%
  \BibitemOpen
  \bibfield  {author} {\bibinfo {author} {\bibfnamefont {M.}~\bibnamefont
  {L{\'o}pez-Su{\'a}rez}}, \bibinfo {author} {\bibfnamefont {I.}~\bibnamefont
  {Neri}}, \ and\ \bibinfo {author} {\bibfnamefont {R.}~\bibnamefont
  {Rurali}},\ }\href@noop {} {\bibfield  {journal} {\bibinfo  {journal}
  {Journal of Applied Physics}\ }\textbf {\bibinfo {volume} {124}} (\bibinfo
  {year} {2018})}\BibitemShut {NoStop}%
\bibitem [{\citenamefont {Xia}\ \emph {et~al.}(2024)\citenamefont {Xia},
  \citenamefont {Wen}, \citenamefont {Chen} \emph {et~al.}}]{xia2024}%
  \BibitemOpen
  \bibfield  {author} {\bibinfo {author} {\bibfnamefont {G.}~\bibnamefont
  {Xia}}, \bibinfo {author} {\bibfnamefont {X.}~\bibnamefont {Wen}}, \bibinfo
  {author} {\bibfnamefont {X.}~\bibnamefont {Chen}},  \emph {et~al.},\
  }\href@noop {} {\bibfield  {journal} {\bibinfo  {journal} {International
  Communications in Heat and Mass Transfer}\ }\textbf {\bibinfo {volume}
  {159}},\ \bibinfo {pages} {108167} (\bibinfo {year} {2024})}\BibitemShut
  {NoStop}%
\bibitem [{\citenamefont {HaiyangLi}\ \emph {et~al.}(2024)\citenamefont
  {HaiyangLi}, \citenamefont {JunWang}, \citenamefont {Xia}, \citenamefont
  {Wen},\ and\ \citenamefont {Chen}}]{HAIYANGLI2024}%
  \BibitemOpen
  \bibfield  {author} {\bibinfo {author} {\bibnamefont {HaiyangLi}}, \bibinfo
  {author} {\bibnamefont {JunWang}}, \bibinfo {author} {\bibfnamefont
  {G.}~\bibnamefont {Xia}}, \bibinfo {author} {\bibfnamefont {X.}~\bibnamefont
  {Wen}}, \ and\ \bibinfo {author} {\bibfnamefont {X.}~\bibnamefont {Chen}},\
  }\href {\doibase https://doi.org/10.1016/j.icheatmasstransfer.2024.108167}
  {\bibfield  {journal} {\bibinfo  {journal} {International Communications in
  Heat and Mass Transfer}\ }\textbf {\bibinfo {volume} {159}},\ \bibinfo
  {pages} {108167} (\bibinfo {year} {2024})}\BibitemShut {NoStop}%
\bibitem [{\citenamefont {Chen}(2021)}]{chen2021}%
  \BibitemOpen
  \bibfield  {author} {\bibinfo {author} {\bibfnamefont {M.}~\bibnamefont
  {Chen}},\ }\href {\doibase 10.1140/epjb/s10051-021-00220-w} {\bibfield
  {journal} {\bibinfo  {journal} {The European Physical Journal B}\ }\textbf
  {\bibinfo {volume} {94}} (\bibinfo {year} {2021}),\
  10.1140/epjb/s10051-021-00220-w}\BibitemShut {NoStop}%
\bibitem [{\citenamefont {Hu}\ \emph {et~al.}(2023)\citenamefont {Hu},
  \citenamefont {Liu}, \citenamefont {Yang}, \citenamefont {Shi}, \citenamefont
  {Antezza}, \citenamefont {Wu},\ and\ \citenamefont {Sun}}]{hu2023}%
  \BibitemOpen
  \bibfield  {author} {\bibinfo {author} {\bibfnamefont {Y.}~\bibnamefont
  {Hu}}, \bibinfo {author} {\bibfnamefont {H.}~\bibnamefont {Liu}}, \bibinfo
  {author} {\bibfnamefont {B.}~\bibnamefont {Yang}}, \bibinfo {author}
  {\bibfnamefont {K.}~\bibnamefont {Shi}}, \bibinfo {author} {\bibfnamefont
  {M.}~\bibnamefont {Antezza}}, \bibinfo {author} {\bibfnamefont
  {X.}~\bibnamefont {Wu}}, \ and\ \bibinfo {author} {\bibfnamefont
  {Y.}~\bibnamefont {Sun}},\ }\href@noop {} {\bibfield  {journal} {\bibinfo
  {journal} {Physical Review Materials}\ }\textbf {\bibinfo {volume} {7}},\
  \bibinfo {pages} {035201} (\bibinfo {year} {2023})}\BibitemShut {NoStop}%
\bibitem [{\citenamefont {Kommandur}\ \emph {et~al.}(2022)\citenamefont
  {Kommandur}, \citenamefont {Kishore}, \citenamefont {Booten}, \citenamefont
  {Cui}, \citenamefont {Wheeler},\ and\ \citenamefont {Vidal}}]{kommandur22}%
  \BibitemOpen
  \bibfield  {author} {\bibinfo {author} {\bibfnamefont {S.}~\bibnamefont
  {Kommandur}}, \bibinfo {author} {\bibfnamefont {R.~A.}\ \bibnamefont
  {Kishore}}, \bibinfo {author} {\bibfnamefont {C.}~\bibnamefont {Booten}},
  \bibinfo {author} {\bibfnamefont {S.}~\bibnamefont {Cui}}, \bibinfo {author}
  {\bibfnamefont {L.~M.}\ \bibnamefont {Wheeler}}, \ and\ \bibinfo {author}
  {\bibfnamefont {J.}~\bibnamefont {Vidal}},\ }\href@noop {} {\bibfield
  {journal} {\bibinfo  {journal} {Advanced Materials Technologies}\ }\textbf
  {\bibinfo {volume} {7}},\ \bibinfo {pages} {2101060} (\bibinfo {year}
  {2022})}\BibitemShut {NoStop}%
\bibitem [{\citenamefont {Avanessian}\ and\ \citenamefont
  {Hwang}(2016)}]{Avanessian2016}%
  \BibitemOpen
  \bibfield  {author} {\bibinfo {author} {\bibfnamefont {T.}~\bibnamefont
  {Avanessian}}\ and\ \bibinfo {author} {\bibfnamefont {G.}~\bibnamefont
  {Hwang}},\ }\href {\doibase 10.1063/1.4966599} {\bibfield  {journal}
  {\bibinfo  {journal} {Journal of Applied Physics}\ }\textbf {\bibinfo
  {volume} {120}} (\bibinfo {year} {2016}),\ 10.1063/1.4966599}\BibitemShut
  {NoStop}%
\bibitem [{\citenamefont {Avanessian}\ and\ \citenamefont
  {Hwang}(2018)}]{AVANESSIAN18}%
  \BibitemOpen
  \bibfield  {author} {\bibinfo {author} {\bibfnamefont {T.}~\bibnamefont
  {Avanessian}}\ and\ \bibinfo {author} {\bibfnamefont {G.}~\bibnamefont
  {Hwang}},\ }\href {\doibase
  https://doi.org/10.1016/j.ijheatmasstransfer.2018.03.039} {\bibfield
  {journal} {\bibinfo  {journal} {International Journal of Heat and Mass
  Transfer}\ }\textbf {\bibinfo {volume} {124}},\ \bibinfo {pages} {201}
  (\bibinfo {year} {2018})}\BibitemShut {NoStop}%
\bibitem [{\citenamefont {Edalatpour}\ \emph {et~al.}(2020)\citenamefont
  {Edalatpour}, \citenamefont {Murphy}, \citenamefont {Mukherjee},\ and\
  \citenamefont {Boreyko}}]{edalatpour20}%
  \BibitemOpen
  \bibfield  {author} {\bibinfo {author} {\bibfnamefont {M.}~\bibnamefont
  {Edalatpour}}, \bibinfo {author} {\bibfnamefont {K.~R.}\ \bibnamefont
  {Murphy}}, \bibinfo {author} {\bibfnamefont {R.}~\bibnamefont {Mukherjee}}, \
  and\ \bibinfo {author} {\bibfnamefont {J.~B.}\ \bibnamefont {Boreyko}},\
  }\href@noop {} {\bibfield  {journal} {\bibinfo  {journal} {Advanced
  Functional Materials}\ }\textbf {\bibinfo {volume} {30}},\ \bibinfo {pages}
  {2004451} (\bibinfo {year} {2020})}\BibitemShut {NoStop}%
\bibitem [{\citenamefont {Asano}\ and\ \citenamefont
  {Fuchizaki}(2012)}]{Asano2012}%
  \BibitemOpen
  \bibfield  {author} {\bibinfo {author} {\bibfnamefont {Y.}~\bibnamefont
  {Asano}}\ and\ \bibinfo {author} {\bibfnamefont {K.}~\bibnamefont
  {Fuchizaki}},\ }\href {\doibase 10.1063/1.4764855} {\bibfield  {journal}
  {\bibinfo  {journal} {The Journal of Chemical Physics}\ }\textbf {\bibinfo
  {volume} {137}},\ \bibinfo {pages} {174502} (\bibinfo {year}
  {2012})}\BibitemShut {NoStop}%
\bibitem [{\citenamefont {Stephan}\ \emph {et~al.}(2018)\citenamefont
  {Stephan}, \citenamefont {Liu}, \citenamefont {Langenbach}, \citenamefont
  {Chapman},\ and\ \citenamefont {Hasse}}]{Stephan_2018}%
  \BibitemOpen
  \bibfield  {author} {\bibinfo {author} {\bibfnamefont {S.}~\bibnamefont
  {Stephan}}, \bibinfo {author} {\bibfnamefont {J.}~\bibnamefont {Liu}},
  \bibinfo {author} {\bibfnamefont {K.}~\bibnamefont {Langenbach}}, \bibinfo
  {author} {\bibfnamefont {W.~G.}\ \bibnamefont {Chapman}}, \ and\ \bibinfo
  {author} {\bibfnamefont {H.}~\bibnamefont {Hasse}},\ }\href {\doibase
  10.1021/acs.jpcc.8b06332} {\bibfield  {journal} {\bibinfo  {journal} {The
  Journal of Physical Chemistry C}\ }\textbf {\bibinfo {volume} {122}},\
  \bibinfo {pages} {24705 } (\bibinfo {year} {2018})}\BibitemShut {NoStop}%
\bibitem [{\citenamefont {Grest}\ and\ \citenamefont
  {Kremer}(1986)}]{Grest_86}%
  \BibitemOpen
  \bibfield  {author} {\bibinfo {author} {\bibfnamefont {G.~S.}\ \bibnamefont
  {Grest}}\ and\ \bibinfo {author} {\bibfnamefont {K.}~\bibnamefont {Kremer}},\
  }\href {\doibase 10.1103/PhysRevA.33.3628} {\bibfield  {journal} {\bibinfo
  {journal} {Phys. Rev. A}\ }\textbf {\bibinfo {volume} {33}},\ \bibinfo
  {pages} {3628} (\bibinfo {year} {1986})}\BibitemShut {NoStop}%
\bibitem [{\citenamefont {Everaers}\ \emph {et~al.}(2020)\citenamefont
  {Everaers}, \citenamefont {Karimi-Varzaneh}, \citenamefont {Fleck},
  \citenamefont {Hojdis},\ and\ \citenamefont {Svaneborg}}]{Everaers2020}%
  \BibitemOpen
  \bibfield  {author} {\bibinfo {author} {\bibfnamefont {R.}~\bibnamefont
  {Everaers}}, \bibinfo {author} {\bibfnamefont {H.~A.}\ \bibnamefont
  {Karimi-Varzaneh}}, \bibinfo {author} {\bibfnamefont {F.}~\bibnamefont
  {Fleck}}, \bibinfo {author} {\bibfnamefont {N.}~\bibnamefont {Hojdis}}, \
  and\ \bibinfo {author} {\bibfnamefont {C.}~\bibnamefont {Svaneborg}},\ }\href
  {\doibase 10.1021/acs.macromol.9b02428} {\bibfield  {journal} {\bibinfo
  {journal} {Macromolecules}\ } (\bibinfo {year} {2020}),\
  10.1021/acs.macromol.9b02428}\BibitemShut {NoStop}%
\bibitem [{\citenamefont {Speyer}\ and\ \citenamefont
  {Pastorino}(2019)}]{Speyer_2019a}%
  \BibitemOpen
  \bibfield  {author} {\bibinfo {author} {\bibfnamefont {K.}~\bibnamefont
  {Speyer}}\ and\ \bibinfo {author} {\bibfnamefont {C.}~\bibnamefont
  {Pastorino}},\ }\href {\doibase 10.1039/c8sm02388c} {\bibfield  {journal}
  {\bibinfo  {journal} {Soft Matter}\ }\textbf {\bibinfo {volume} {15}},\
  \bibinfo {pages} {937} (\bibinfo {year} {2019})}\BibitemShut {NoStop}%
\bibitem [{\citenamefont {Speyer}\ and\ \citenamefont
  {Pastorino}(2017)}]{Speyer_17}%
  \BibitemOpen
  \bibfield  {author} {\bibinfo {author} {\bibfnamefont {K.}~\bibnamefont
  {Speyer}}\ and\ \bibinfo {author} {\bibfnamefont {C.}~\bibnamefont
  {Pastorino}},\ }\href {\doibase 10.1021/acs.langmuir.7b02640} {\bibfield
  {journal} {\bibinfo  {journal} {Langmuir}\ }\textbf {\bibinfo {volume}
  {33}},\ \bibinfo {pages} {10753} (\bibinfo {year} {2017})}\BibitemShut
  {NoStop}%
\bibitem [{\citenamefont {Speyer}\ and\ \citenamefont
  {Pastorino}(2015)}]{Speyer_15}%
  \BibitemOpen
  \bibfield  {author} {\bibinfo {author} {\bibfnamefont {K.}~\bibnamefont
  {Speyer}}\ and\ \bibinfo {author} {\bibfnamefont {C.}~\bibnamefont
  {Pastorino}},\ }\href {\doibase 10.1039/C5SM01075F} {\bibfield  {journal}
  {\bibinfo  {journal} {Soft Matter}\ }\textbf {\bibinfo {volume} {11}},\
  \bibinfo {pages} {5473} (\bibinfo {year} {2015})}\BibitemShut {NoStop}%
\bibitem [{\citenamefont {M{\"u}ller}\ and\ \citenamefont
  {MacDowell}(2000)}]{Mueller_00}%
  \BibitemOpen
  \bibfield  {author} {\bibinfo {author} {\bibfnamefont {M.}~\bibnamefont
  {M{\"u}ller}}\ and\ \bibinfo {author} {\bibfnamefont {L.~G.}\ \bibnamefont
  {MacDowell}},\ }\href {\doibase 10.1021/ma991796t} {\bibfield  {journal}
  {\bibinfo  {journal} {Macromolecules}\ }\textbf {\bibinfo {volume} {33}},\
  \bibinfo {pages} {3902} (\bibinfo {year} {2000})},\ \Eprint
  {http://arxiv.org/abs/http://dx.doi.org/10.1021/ma991796t}
  {http://dx.doi.org/10.1021/ma991796t} \BibitemShut {NoStop}%
\bibitem [{\citenamefont {Pastorino}\ \emph {et~al.}(2006)\citenamefont
  {Pastorino}, \citenamefont {Binder}, \citenamefont {Kreer},\ and\
  \citenamefont {M{\"u}ller}}]{Pastorino_06}%
  \BibitemOpen
  \bibfield  {author} {\bibinfo {author} {\bibfnamefont {C.}~\bibnamefont
  {Pastorino}}, \bibinfo {author} {\bibfnamefont {K.}~\bibnamefont {Binder}},
  \bibinfo {author} {\bibfnamefont {T.}~\bibnamefont {Kreer}}, \ and\ \bibinfo
  {author} {\bibfnamefont {M.}~\bibnamefont {M{\"u}ller}},\ }\href@noop {}
  {\bibfield  {journal} {\bibinfo  {journal} {J. Comp. Phys.}\ }\textbf
  {\bibinfo {volume} {124}},\ \bibinfo {pages} {064902} (\bibinfo {year}
  {2006})}\BibitemShut {NoStop}%
\bibitem [{\citenamefont {Tehver}\ \emph {et~al.}(1998)\citenamefont {Tehver},
  \citenamefont {Toigo}, \citenamefont {Koplik},\ and\ \citenamefont
  {Banavar}}]{Tehver_98}%
  \BibitemOpen
  \bibfield  {author} {\bibinfo {author} {\bibfnamefont {R.}~\bibnamefont
  {Tehver}}, \bibinfo {author} {\bibfnamefont {F.}~\bibnamefont {Toigo}},
  \bibinfo {author} {\bibfnamefont {J.}~\bibnamefont {Koplik}}, \ and\ \bibinfo
  {author} {\bibfnamefont {J.~R.}\ \bibnamefont {Banavar}},\ }\href {\doibase
  10.1103/PhysRevE.57.R17} {\bibfield  {journal} {\bibinfo  {journal} {Phys.
  Rev. E}\ }\textbf {\bibinfo {volume} {57}},\ \bibinfo {pages} {17} (\bibinfo
  {year} {1998})}\BibitemShut {NoStop}%
\bibitem [{\citenamefont {Marsaglia}\ and\ \citenamefont
  {Tsang}(2000)}]{Marsaglia2000}%
  \BibitemOpen
  \bibfield  {author} {\bibinfo {author} {\bibfnamefont {G.}~\bibnamefont
  {Marsaglia}}\ and\ \bibinfo {author} {\bibfnamefont {W.~W.}\ \bibnamefont
  {Tsang}},\ }\href {\doibase 10.18637/jss.v005.i08} {\bibfield  {journal}
  {\bibinfo  {journal} {Journal of Statistical Software}\ }\textbf {\bibinfo
  {volume} {5}} (\bibinfo {year} {2000}),\ 10.18637/jss.v005.i08}\BibitemShut
  {NoStop}%
\bibitem [{\citenamefont {Urrutia}\ and\ \citenamefont
  {Pastorino}(2014)}]{Urrutia_14b}%
  \BibitemOpen
  \bibfield  {author} {\bibinfo {author} {\bibfnamefont {I.}~\bibnamefont
  {Urrutia}}\ and\ \bibinfo {author} {\bibfnamefont {C.}~\bibnamefont
  {Pastorino}},\ }\href {\doibase 10.1063/1.4896221} {\bibfield  {journal}
  {\bibinfo  {journal} {The Journal of Chemical Physics}\ }\textbf {\bibinfo
  {volume} {141}},\ \bibinfo {pages} {124905} (\bibinfo {year} {2014})},\
  \Eprint {http://arxiv.org/abs/https://doi.org/10.1063/1.4896221}
  {https://doi.org/10.1063/1.4896221} \BibitemShut {NoStop}%
\bibitem [{\citenamefont {Paganini}\ \emph {et~al.}(2015)\citenamefont
  {Paganini}, \citenamefont {Pastorino},\ and\ \citenamefont
  {Urrutia}}]{Paganini_15}%
  \BibitemOpen
  \bibfield  {author} {\bibinfo {author} {\bibfnamefont {I.~E.}\ \bibnamefont
  {Paganini}}, \bibinfo {author} {\bibfnamefont {C.}~\bibnamefont {Pastorino}},
  \ and\ \bibinfo {author} {\bibfnamefont {I.}~\bibnamefont {Urrutia}},\ }\href
  {\doibase 10.1063/1.4923164} {\bibfield  {journal} {\bibinfo  {journal} {The
  Journal of Chemical Physics}\ }\textbf {\bibinfo {volume} {142}},\ \bibinfo
  {pages} {244707} (\bibinfo {year} {2015})},\ \Eprint
  {http://arxiv.org/abs/http://dx.doi.org/10.1063/1.4923164}
  {http://dx.doi.org/10.1063/1.4923164} \BibitemShut {NoStop}%
\bibitem [{\citenamefont {Pastorino}\ \emph {et~al.}(2022)\citenamefont
  {Pastorino}, \citenamefont {Urrutia}, \citenamefont {Fiora},\ and\
  \citenamefont {Condado}}]{Pastorino2022}%
  \BibitemOpen
  \bibfield  {author} {\bibinfo {author} {\bibfnamefont {C.}~\bibnamefont
  {Pastorino}}, \bibinfo {author} {\bibfnamefont {I.}~\bibnamefont {Urrutia}},
  \bibinfo {author} {\bibfnamefont {M.}~\bibnamefont {Fiora}}, \ and\ \bibinfo
  {author} {\bibfnamefont {F.}~\bibnamefont {Condado}},\ }\href {\doibase
  10.1088/1361-648x/ac77ce} {\bibfield  {journal} {\bibinfo  {journal} {Journal
  of Physics: Condensed Matter}\ }\textbf {\bibinfo {volume} {34}},\ \bibinfo
  {pages} {344004} (\bibinfo {year} {2022})}\BibitemShut {NoStop}%
\bibitem [{\citenamefont {Smith}\ \emph {et~al.}(2019)\citenamefont {Smith},
  \citenamefont {Daivis},\ and\ \citenamefont {Todd}}]{Smith_2019}%
  \BibitemOpen
  \bibfield  {author} {\bibinfo {author} {\bibfnamefont {E.~R.}\ \bibnamefont
  {Smith}}, \bibinfo {author} {\bibfnamefont {P.~J.}\ \bibnamefont {Daivis}}, \
  and\ \bibinfo {author} {\bibfnamefont {B.~D.}\ \bibnamefont {Todd}},\ }\href
  {\doibase 10.1063/1.5079993} {\bibfield  {journal} {\bibinfo  {journal} {The
  Journal of Chemical Physics}\ }\textbf {\bibinfo {volume} {150}},\ \bibinfo
  {pages} {064103} (\bibinfo {year} {2019})}\BibitemShut {NoStop}%
\bibitem [{\citenamefont {Bugel}\ and\ \citenamefont
  {Galliero}(2008)}]{Bugel2008}%
  \BibitemOpen
  \bibfield  {author} {\bibinfo {author} {\bibfnamefont {M.}~\bibnamefont
  {Bugel}}\ and\ \bibinfo {author} {\bibfnamefont {G.}~\bibnamefont
  {Galliero}},\ }\href {\doibase 10.1016/j.chemphys.2008.06.013} {\bibfield
  {journal} {\bibinfo  {journal} {Chemical Physics}\ }\textbf {\bibinfo
  {volume} {352}},\ \bibinfo {pages} {249} (\bibinfo {year}
  {2008})}\BibitemShut {NoStop}%
\bibitem [{\citenamefont {Henry}\ and\ \citenamefont {Chen}(2008)}]{Henry_08}%
  \BibitemOpen
  \bibfield  {author} {\bibinfo {author} {\bibfnamefont {A.}~\bibnamefont
  {Henry}}\ and\ \bibinfo {author} {\bibfnamefont {G.}~\bibnamefont {Chen}},\
  }\href {\doibase 10.1103/physrevlett.101.235502} {\bibfield  {journal}
  {\bibinfo  {journal} {Physical Review Letters}\ }\textbf {\bibinfo {volume}
  {101}} (\bibinfo {year} {2008}),\ 10.1103/physrevlett.101.235502}\BibitemShut
  {NoStop}%
\bibitem [{\citenamefont {Henry}(2014)}]{Henry2014}%
  \BibitemOpen
  \bibfield  {author} {\bibinfo {author} {\bibfnamefont {A.}~\bibnamefont
  {Henry}},\ }\href {\doibase 10.1615/annualrevheattransfer.2013006949}
  {\bibfield  {journal} {\bibinfo  {journal} {Annual Review of Heat Transfer}\
  }\textbf {\bibinfo {volume} {17}},\ \bibinfo {pages} {485} (\bibinfo {year}
  {2014})}\BibitemShut {NoStop}%
\bibitem [{\citenamefont {Liu}\ and\ \citenamefont {Yang}(2012)}]{Liu2012}%
  \BibitemOpen
  \bibfield  {author} {\bibinfo {author} {\bibfnamefont {J.}~\bibnamefont
  {Liu}}\ and\ \bibinfo {author} {\bibfnamefont {R.}~\bibnamefont {Yang}},\
  }\href {\doibase 10.1103/physrevb.86.104307} {\bibfield  {journal} {\bibinfo
  {journal} {Physical Review B}\ }\textbf {\bibinfo {volume} {86}},\ \bibinfo
  {pages} {104307} (\bibinfo {year} {2012})}\BibitemShut {NoStop}%
\bibitem [{\citenamefont {Zhao}\ \emph {et~al.}(2023)\citenamefont {Zhao},
  \citenamefont {Wu}, \citenamefont {Sun}, \citenamefont {Lin}, \citenamefont
  {Zhong}, \citenamefont {Jiang},\ and\ \citenamefont {Wu}}]{ZHAO23}%
  \BibitemOpen
  \bibfield  {author} {\bibinfo {author} {\bibfnamefont {H.}~\bibnamefont
  {Zhao}}, \bibinfo {author} {\bibfnamefont {Y.}~\bibnamefont {Wu}}, \bibinfo
  {author} {\bibfnamefont {H.}~\bibnamefont {Sun}}, \bibinfo {author}
  {\bibfnamefont {B.}~\bibnamefont {Lin}}, \bibinfo {author} {\bibfnamefont
  {M.}~\bibnamefont {Zhong}}, \bibinfo {author} {\bibfnamefont
  {G.}~\bibnamefont {Jiang}}, \ and\ \bibinfo {author} {\bibfnamefont
  {S.}~\bibnamefont {Wu}},\ }\href {\doibase
  https://doi.org/10.1016/j.renene.2023.119278} {\bibfield  {journal} {\bibinfo
   {journal} {Renewable Energy}\ }\textbf {\bibinfo {volume} {218}},\ \bibinfo
  {pages} {119278} (\bibinfo {year} {2023})}\BibitemShut {NoStop}%
\end{thebibliography}

%

\end{document}